%% file: Journal_main.tex
\documentclass[journal,comsoc]{IEEEtran}
\usepackage[caption=false,font=footnotesize]{subfig}

\usepackage[T1]{fontenc}
\newtheorem{thm}{Theorem}
\newtheorem{lemma}{Lemma}
\newtheorem{proposition}{Proposition}
\newtheorem{cor}{Corollary}

\newcommand{\CPi}{\mbox{CP}_i}
\newcommand{\UCPi}{U_{\mbox{\tiny CP}_i}}
\newcommand{\UISP}{U_{\mbox{\tiny ISP}}}
\newcommand{\N}{\mathcal{N}}

\usepackage{amssymb,amsbsy,epsfig,float,bm}
\usepackage{amsmath}

\usepackage{ifthen}
\usepackage{color}
\newboolean{showcomments}
\setboolean{showcomments}{true}
\newcommand{\mh}[1]{  \ifthenelse{\boolean{showcomments}}
{ \textcolor{red}{(MH says:  #1)}} {}  }
\newcommand{\jk}[1]{  \ifthenelse{\boolean{showcomments}}
{ \textcolor{blue}{(JK says:  #1)}} {}  }
\newcommand{\fm}[1]{  \ifthenelse{\boolean{showcomments}}
{ \textcolor{green}{(FM says:  #1)}} {}  }
\newcommand{\yh}[1]{  \ifthenelse{\boolean{showcomments}} 
{ \textcolor{red}{(YH says:  #1)}} {}  }

\usepackage[cmintegrals]{newtxmath}

\begin{document}
	
	\title{Revenue Sharing in the Internet: A Moral Hazard Approach and a Net-neutrality Perspective}

	\author{Fehmina Malik, Manjesh K.~Hanawal,
		Yezekael Hayel and Jayakrishnan Nair
		\IEEEcompsocitemizethanks{\IEEEcompsocthanksitem Fehmina Malik and Manjesh K. Hanawal are with IEOR, IIT Bombay, India. E-mail: \{fehminam, mhanawal, \}@iitb.ac.in. Yezekael Hayel is with LIA/CERI, University of Avignon, France. E-mail: yezekael.hayel@univ-avignon.fr.  Jayakrishnan Nair is with Department of Electrical Engineering, IIT Bombay, E-mail: jayakrishnan.nair@ee.iitb.ac.in
		}
	}

	\maketitle
	
	\begin{abstract}
		Revenue sharing contracts between Content Providers (CPs) and Internet Service Providers (ISPs) can act as leverage for enhancing the infrastructure of the Internet. ISPs can be incentivised to make investments in network infrastructure that improve Quality of Service (QoS) for users if attractive contracts are negotiated between them and CPs. The idea here is that part of the net profit gained by CPs are given to ISPs to invest in the network. The Moral Hazard economic framework is used to model such an interaction, in which a principal determines a contract, and an agent reacts by adapting her effort. In our setting, several competitive CPs interact through one common ISP. Two cases are studied: (i) the ISP differentiates between the CPs and makes a (potentially) different investment to improve the QoS of each CP, and (ii) the ISP does not differentiate between CPs and makes a common investment for both.  The last scenario can be viewed as \emph{network neutral behavior} on the part of the ISP. We analyse the optimal contracts and show that the CP that can better monetize its demand always prefers the non-neutral regime.
		Interestingly, ISP revenue, as well as social utility, are also found
		to be higher under the non-neutral regime.
	\end{abstract}
	

	\section{Introduction}
	\input{introduction}

	\section{Problem Description}
	\label{sec:problem}
	\input{problem}

	\section{Symmetric Case}
	\label{sec:MultipleSymmetric}
	\input{MultiSymmetry}
	\section{Asymmetric case}
	\label{sec:AsymetricCase}
	\input{AsymmetricCase}
	
	\section{Conclusions and Regulatory Issues}   
	\label{sec:Conlcusion}
	
	We studied the problem of revenue sharing between multiple CPs and an ISP on the Internet using the moral hazard  framework  with multiple principles and a single agent. We compared the revenues of each player and the social utility in a regime where the ISP is forced to put equal effort for all the CPs (neutral) with a regime where there are no such restrictions (non-neutral) on the ISP. Our key take-away is that every one is better off and social utility is higher in non-neutral regime when the CPs ability to montize their demand are `nearly' the same. When the there is a large disparity in the monetization power of the CPs, for the case of two CPs we showed that non-neutral regime is preferable from the standpoint of the dominant CP (with higher monetizing power), the ISP,
	and from the standpoint of social utility. On the other hand, the non-dominant CP is benefited by a neutrality stipulation since it gets to `free-ride' on the contribution made by the dominant CP, the very reason that makes this regime less preferred by the dominat CP. 
	
	Our analysis throws up an intriguing dilemma for a regulator --enforcing neutrality brings in parity in the way ISP treat the CPs, but it worsens the social utility and pay-off of all the players compared to the neutral regime if the players act non-cooperatively. It is then interesting to study mechanisms that the regulator can use to induce cooperation among the the players so that the social utility and players pay-off are no worse than in the non-neutral regime.

	\appendix
	\input{Appendix}
	\bibliographystyle{IEEEtran}	
	\bibliography{bib}

\end{document}

%% file: introduction.tex
The rapid growth of data-intensive services has resulted in an explosion of the Internet traffic, and it is expected to increase at an even faster rate in the future \cite{jung2011cisco}. 
To accommodate this increase in traffic, and to provide better Quality of Service (QoS) for end users, Internet service providers (ISPs) need to upgrade their network infrastructure and expand capacity. This development follows the deployment of next generation networks that will induce more business interactions between service providers and content providers. For example, caching technologies has recently received increased attention from industry and academia to be a key solution for next generation networks \cite{Paschos2018jsac}. As it is specified in this special issue, the economics of caching will be one of important aspect in deciding the monetary interactions between the ISPs and CPs.

In terms of return on investments, the ISPs, especially the ones providing last-mile connectivity, feel that the revenue from end-users (mainly access charges) are often not enough to recoup their investment costs and they propose that the CPs share the risk by sharing part of their revenues. The CPs may also have the incentive to contribute to ISP capacity expansion, as increased capacity and better QoS trigger higher demand for content and help them earn even higher revenues (mainly from subscriptions and advertisements) \cite{sen2013survey}. For example, in \cite{Krolikowski2018jsac}, the authors propose a model in which a Mobile  Network  Operator  leases its  edge caches to  a Content Provider. This is an increasingly relevant scenario and follows proposals for deploying edge storage resources at mobile 5G networks \cite{Bastug2014ieeemag}. A recent announcement by Comcast that it will bundle Netflix subscription in its package is another example of an ISP helping a CP to increase its demand.



If a CP enters into a contract with an ISP to share its revenue in return for ISP putting efforts to improve demand for its content, the CP may want to monitor the efforts level of ISP so that the contract is honored.  However such monitoring of efforts levels may not be always feasible in the Internet as there is an inherent asymmetry of information between CPs and ISPs -- CPs cannot observe the exact investment (caching effort) made by the ISP, but can only observe the resulting increase in user demand for its content, which is a (random) function of the ISP's investment. In this situation, an obvious question arises: how can this imperfect information about the ISP's investment be used to formulate an optimal revenue sharing contract?  This is a situation where a privately taken action (investment) by the ISP influences the probability distribution of the outcome (demand) for the CP.

The role of this information asymmetry between the CP and ISP is critical under such agreements, where ISP makes an investment which ultimately benefits the CP, and should be considered in the structure of the contract. Therefore, moral hazard can be applied to propose a contract in which the ISP (agent) knows that CP (principle) will pay to cover its risks, which in turn gives the ISP the incentive to make the (risky) investment. Also, the moral hazard model is proven to induce proper incentives for taking appropriate action \cite{holmstrom1979moral}, and this may provide fair revenue sharing between CPs and ISPs.  In our work, we propose an incentivizing mechanism using the \textit{moral hazard} approach in which CPs share a part of their revenues with an ISP expecting a better QoS for their content and higher revenue in return from the ISP efforts.

The classical moral hazard problem deals with a single principle and a single agent, whereas we are faced with possibility of multiple principles interacting with multiple agents. In this work, we focus on a monopolistic ISP connecting end users to multiple CPs, i.e., multiple principles and a single agent. Our interest is in determining optimal sharing mechanisms/contracts between each CP and the ISP. We distinguish two cases for the effort (investment) made by the ISP. In one case, we allow the ISP to make different amounts of effort for each CP, and in the other case, the ISP is constrained to put an equal amount of effort for all CPs. The former case corresponds to a `non-neutral' regime where the ISP is allowed to differentiate between CPs, and the latter case corresponds to a `neutral' regime where the ISP cannot differentiate between CPs. We compare the revenue of each players and social utilities under both the regimes and analyze which regime is preferred by the players. 

We consider competition between multiple CPs that provide content of a
particular type, for example, video, and aim to earn higher revenue by
entering into a revenue sharing contract with the ISP. Each CP
separately negotiates with the ISP knowing that the other CPs can also
enter into similar negotiations with the ISP. We consider linear contracts and analyze the
equilibrium sharing contracts in both the neutral and non-neutral
regime. We first consider the symmetric case where all the CPs earn same revenue per unit demand. This corresponds to the case where ability of the CPs to monitize their demand is the same. We then study the asymetric case where the CPs capability to monetize their demand could be different. Our
contributions and observations can be summarized as follows:
\begin{itemize}
\item We model the competitive revenue sharing of CPs with an ISP in return for improved QoS in the moral hazard framework with multiple principles and a single agent.
\item We analyze the equilibrium contracts in a regime where the ISP
can put a different level of effort for each CP (non-neutral) and in a
regime where it is constrained to put equal efforts for all CPs
(neutral).
\item In the symmetric case we show that all the players prefer the non-neutral regimes as their utilities are higher
\item In the asymmetric case we show that the CP which can better monetize the demand for
its traffic always prefers the non-neutral regime whereas the CP with weaker monetization power can prefer the neutral or non-neutral regime depending on its relative monetization power.
\item The ISP always prefers the non-neutral regime. Moreover,
social utility (defined as the sum of the average earning of all
players) is also higher in the non-neutral regime.
\end{itemize}

This paper is organized as follows. In Section~\ref{sec:problem} we
discuss the problem setup and define contracts under the neutral and non-neutral regime. We study the equilibrium contracts under neutral and non-neutral regime under the symmetric case in in Section~\ref{sec:MultipleSymmetric} and under asymetric case in Section~\ref{sec:AsymetricCase}. Conclusions and future extensions are discussed in Section~\ref{sec:Conlcusion}. Proofs of all stated results can be
found in the appendix.

\subsection{Related works}
Several works \cite{kamiyama2014effect, kamiyama2014feasibility,
park2014isp1, park2014isp2, im2016revenue} study the possibility of
content charges by ISPs to recover investment
costs. In \cite{kamiyama2014effect}
and \cite{kamiyama2014feasibility}, the authors investigated the
feasibility of ISPs charging a content charge to CPs, and evaluate its
effect by modeling the Stackelberg game between CPs and
ISPs. In \cite{park2014isp1} and \cite{park2014isp2}, a revenue-sharing scheme is proposed when the ISP provides a content piracy
monitoring service to CPs for increasing the demand for their content.
This work is extended to two ISPs competing with each other
in \cite{im2016revenue} where only one of them provides the content
piracy monitoring service. Several studies considered
cooperative settlement between service providers for profit
sharing \cite{ma2010,ma2011,yeze2017} where the mechanisms are derived
using the Shapely value concept.


A moral hazard framework is applied in \cite{constantiou2001information} to study interconnection contracts between ISP and end users in the market for network transport services. However, the contract design problem between ISPs and CPs remained unaddressed in this work. In the present paper, we apply the moral hazard approach to study revenue sharing between an ISP and multiple CPs. The CPs act as the principles that enter into a contract with the ISP (agent) to improve QoS for their content. We thus end up with a multiple principle, single agent problem.

A moral hazard framework where two principals offer a contract to the same agent is studied in \cite{commonagency} where the principals can only observe correlated noisy signals of the common one-dimensional action taken by the agent, where the principals' output is linearly increasing with the agent's action. Our work is different from them as we consider that even though the agent (ISP) is common, it chooses different actions for each principal (CP). We also compare it with the setting when the ISP is forced to choose the same action for both CPs.  Also, we consider the demands of CPs to be logarithmic in the efforts of the agent, unlike the linear case considered in \cite{commonagency}. Such a demand function comes from the fact that the delay experienced by end users is exponential in the cache and not linear as assumed in \cite{yeze2017} and \cite{kesidis13}.

A moral hazard setup is also used in \cite{Zhang17} for motivating end users to participate in crowd-sourced services in one principal and several agent problem.  


%% file: problem.tex
We consider multiple Content Providers (CPs) and a single Internet Service Provider (ISP) that connects end users to the content of the CPs. Each CP can enter into a contract with the ISP under which the ISP agrees to offer a better quality of service on the contents of the CPs to the end users by investing in the network infrastructure \cite{caching} and in turn, each CP agrees to share a part of its revenue with the ISP. One example of ISP investing in network infrastructure to improve QoS is caching, where better caching efforts by the ISP for a CP's content results in higher revenue for the CP. However, such caching decisions (effort or action) of ISP may not be directly visible to the CPs, but each CP can observe the QoS experienced by end users through demand for their contents. Thus, higher the effort (caching) by the ISP, higher will be the revenue for CP because of the price paid for their content or from advertisements (from the click-through rate) \cite{Courcou13}. However, the ISP's profit maximization strategy may not be aligned with the interests of the CPs, and moreover, the ISP effort (that influence CP revenue) is not directly observable by CPs.  This scenario gives rise to the Moral Hazard problem.

\begin{figure}[h!]
 	\begin{center}
 		\includegraphics[scale=0.4]{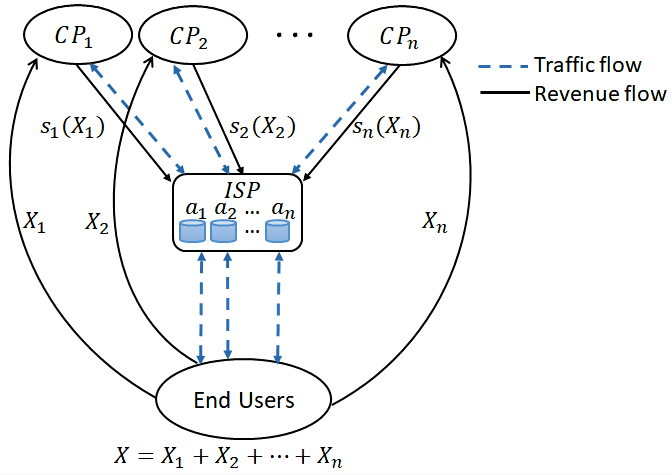}
 		\caption{Revenue flow between CPs, ISP and End Users.}
	\label{fig:figure1}      
\end{center}
\end{figure}
 
Let $n$ denote the number of CPs and $\mathcal{N}=\{1,2,\dots, n\}$ the set of CPs.  For each $i \in N$ we denote $i$-th CP as $\CPi$. 
The revenue of $\CPi$ from its content is  random and denoted as $X_i\in \mathcal{R}_+$ with probability density function $F_i$. The amount of efforts put by the ISP to improve the demand for the content of $\CPi$ is quantified by a positive number denoted as $a_i \in \mathcal{R}_+$. 
$\CPi$ shares part of his revenue with the ISP to incentivize ISP investments. This share is determined by the outcome as sharing function $s_i:\mathcal{R}_+ \rightarrow \mathcal{R}_+$, which is called contract or agreement in the moral hazard framework. Specifically, if $x_i$ is the realized revenue in a month for $\CPi$, it gives  $s_i(x_i)$ to the ISP. Then, the net revenue of $\CPi$ is $x_i-s_i(x_i)$. Our goal in this work is first to design optimal sharing functions $s_i(\cdot), i\in \mathcal{N}$ that maximize the expected net revenues of each CPs taking into account the rational behavior of the ISP and its participation constraints. We assume that the CPs offer a similar type of content and compete with each other to attract more demand.  Our model with multiple CPs and a single ISP is depicted in Figure~\ref{fig:figure1}. In the terminology of moral hazard, the CPs are the principles, and the ISP is the agent. We thus have multiple principles and a single agent.  

\noindent
{\bf Demand and Revenue:} The increment in demand for content of $\CPi$ depends on the effort by the ISP to improve QoS for  $\CPi$ content and could be random. Let $D_i$ denote this increment in demand. We assume that the mean of $D_i$ grows logarithmically in $a_i$ (law of diminishing gains) and is given by $D_i:=\log(a_i+1)+\epsilon_i, i\in \mathcal{N}$,  where $\epsilon_i$ is the random variation in the demand for $\CPi$. 

The revenue generated for $\CPi$ is proportional to the demand and
given by $X_i=r_iD_i$, where $r_i$ is a constant that captures how
each unit of demand translates to earnings. For example, $r_i$ could
be revenue per click for $\CPi$ when $D_i$ is interpreted as total
number of clicks on $\CPi$'s content. The expected revenue for $\CPi$
is then $\mathbb{E}[X_i]=r_i\log(r_i+1)$.\footnote{Our model
generalizes trivially to the case where $D_i:=
d_i \log(a_i+1)+\epsilon_i,$ i.e., the average demand scales with a
CP-specific multiplicative constant; this constant can simply be
absorbed into $r_i.$}


\subsection{Utilities and Objective}
For a given contract $s_i(\cdot), i\in \mathcal{N}$ the net revenue for $CP_i$ is
$X_i-s_i(X_i)$. Let $V_i(\cdot)\in\mathcal{R}$ denote the utility
function of $\CPi$ and is typically assumed to an increasing concave
function of the $CP$'s net revenue. We assume the utility of $CP_i$ is
linear in its net revenue and set
$V_i(X_i-s_i(X_i))=X_i-s_i(X_i)$. The utility of the ISP depends on
revenue-share it receives from both the CPs and also the cost involved
in the efforts it puts for the CPs. The earning for the ISP from the
CPs is $\sum_{i} s_i(X_i)$ while it incurs a total cost of
$c\sum_{i}a_i$, where $c$ is a positive constant. The net earnings for the ISP
is then $W(s_1(X_1) \ldots,s_n(X_n),a_1\ldots, a_n)=\sum_{i}
s_i(X_i)-c\sum_{i}a_i$. We assume the ISP is risk-averse and set its
utility as $H(W)=-\exp\{-zW\}$, where $z$ is a risk-averse
parameter. This utility is referred to as Constant Absolute Risk
Aversion (CARA) in the literature \cite{rabin2013risk}. The ISP enters
into agreement with the CPs only if its expected utility is larger
than a certain threshold denoted as $\overline{H}$.

The CPs compete against each other and aim to maximize their utility. The objective of $\CPi, i\in \mathcal{N}$ taking into participation constraints of the ISP is given as follows:
\begin{eqnarray} \nonumber
&\max_{a_1,...,a_n, s_i(\cdot)}\UCPi:=\mathbb{E}_{X_i}[V_i(X_i-s_i(X_i))]&\\
&\hspace{-5mm}\text{subjected to} \hspace{1mm} \mathbb{E}_{X_1,...,X_n}[H(s_1(X_1),...,s_n(X_n),a_1,...,a_n)]\geq \overline{H}& \label{eqn:IR}\\ 
&\max_{a_1,\ldots,a_2}\mathbb{E}_{X_1,...,X_n}[H(s_1(X_1),...,s_n(X_n),a_1,...,a_n)]. &  \label{eqn:IC}
\end{eqnarray}
Constraint (\ref{eqn:IR}) guarantees the ISP minimum expected utility $\overline{H}$ and is called as individual rationality (IR) or the participation constraint. The second constraint (\ref{eqn:IC}) is the ISP's optimization problem and is called an incentive compatibility constraint (IC). It also captures the fact that the CPs cannot observe the effort level of the ISP. Note that the Constraint in (\ref{eqn:IC}) may have multiple optima hence objective of each CP is optimized over all of these possibilities.  The structure of the optimization problem is hierarchical and can be studied considering a Stackelberg solution concept where the principals (here each CP) can be seen as leaders and the agent (here the ISP) as the follower who plays after observing the action of the principals. In our setting, there is another level of complexity as there are multiple strategic principals. This gives rise to a static game between CPs with shared constraints.



\subsection{Linear Contracts}
In the following we consider a specific type of contract where CPs
share a fraction of their revenue to ISP. These contracts are of the form
$s_i(X_i)=\beta_i X_i$ and are referred to as linear contracts, where
$\beta_i\in [0,1]$ for all $i\in \mathcal{N}$. The ISP chooses whether to accept
or reject the contract. Linear contracts are shown to be optimal
in \cite{holmstrom1991multitask} particularly when agent has a risk
averse utility which is also the case in our setting. Henceforth we denote a linear contract between the ISP and $\CPi$ by its parameter
$\beta_i$.  The expected utility of $CP_i$ is then:
\begin{align*}
\UCPi =\mathbb{E}[X_i(1-\beta_i)]=(1-\beta_i)r_i\log(a_i+1),
\end{align*}
and the expected utility of the $ISP$ is:
\begin{align*}
&\UISP=\mathbb{E}\left [H\left(\sum_{i} (s_i(X_i)-ca_i)\right)\right] \\
&=E\left [H\left (\sum_i(\beta_ir_i(\log(a_i+1)+\epsilon_i)-ca_i)\right)\right]\\
&=-\exp\left\{-z\sum_i(\beta_ir_i\log(a_i+1)-ca_i)\right\} \mathbb{E}\left[\exp\left\{-z\sum_i\beta_ir_i\epsilon_i\right\}\right].
\end{align*}
The ISP's optimization problem is to find an effort level $a_i$ for each CP$_i$ which maximizes its expected
utility. Notice that maximizing expected utility (IC constraint) is
equivalent to maximizing $\sum_i(\beta_ir_i\log(a_i+1)-ca_i)$ over
$(a_1,\ldots, a_n)$.

\subsection{Neutral vs Non-Neutral regime}
We distinguish two scenarios based on differentiation in the efforts by the ISP for each CP. We say that the network is {\em neutral} if ISP puts the same amount of efforts for both the CPs irrespective of the revenue-share it can get from them, i.e., ISP always sets $a_1=a_2...=a_n$. We say that the network is {\em non-neutral} if the ISP can put different amount of effort for each CP, i.e., $a_1\neq a_2...\ne a_n$ is permitted. Hence in the neutral regime the ISP treats each CP identically, whereas it can differentiate between them in the non-neutral regime. 

Under the neutral regime and linear contracts, the IC constraint of the ISP, i.e., $\max_{a}E[H(s_1(X_1),...,s_n(X_n),a)]$, simplifies  to $\max_a \sum_i(\beta_ir_i\log(a+1)-ca)$. The optimal effort level for a given contracts $(\beta_i, i\in \mathcal{N})$ is given by:
	\begin{equation}
a^*=\max\left(\frac{\sum_i\beta_ir_i}{nc}-1,0\right).
\end{equation}
The objective function of $\CPi, i\in \mathcal{N}
$ in the neutral regime can then be expressed as follows:\\
\vspace{.2cm}
\textbf{Neutral:}
\vspace{-.2cm}
\begin{equation}
\label{eqn:OptNeutral}
\begin{aligned}
 &\max_{\beta_i \in [0,1]} \hspace{2mm} (1-\beta_i)\log(a+1)r_i\\
&\text{subjected to} \sum_iz(\beta_ir_i\log(a+1)-ca)\\
&-\log\mathbb{E}\left[\exp\left\{-z\sum_i\beta_ir_i\epsilon_i\right\}\right]\geq -\log (-\overline{H})\\
& \text{ and } a=\max\left(\frac{\sum_i\beta_ir_i}{nc}-1,0\right).\\
\end{aligned} 
\end{equation}
Under the non-neutral regime, the IC constraint of the ISP, i.e., $\max_{a_1,\ldots, a_n}E[H(s_1(X_1),...,s_n(X_n),a_1,...,a_n)]$, simplifies  to $\max_{a_1,\ldots,a_n} \sum_i(\beta_ir_i\log(a_i+1)-ca_i)$.  The optimial efforts level for a given contracts $(\beta_i, i\in \mathcal{N})$ are:
$$a_i^*=\max\left(\frac{\beta_ir_i}{c}-1,0\right) \mbox{ for }  i\in \mathcal{N}.$$
Simplified optimization problem for  each $CP_i$ can then be expressed as the following bi-level optimization problem: \\
\vspace{.2cm}
\textbf{NonNeutral:}
\vspace{-.2cm}
\begin{equation}
\label{eqn:OptNonNeutral}
\begin{aligned}
&\max_{\beta_i \in [0,1]} \hspace{2mm} (1-\beta_i)\log(a_i+1)r_i\\
&\text{subjected to} \sum_i z(\beta_ir_i\log(a_i+1)-ca_i)\\
&-\log\mathbb{E}\left[\exp\left\{-z\sum_i\beta_ir_i\epsilon_i\right\}\right]\geq -\log (-\overline{H})\\
&\text{ and } a_i=\max\left(\frac{\beta_ir_i}{c}-1,0\right) \mbox{ for } i \in \mathcal{N}.
\end{aligned}
\end{equation} 
Notice that the ISP has incentive to enter into contract with the CPs only when $\overline{H}\geq -1$, otherwise its net earning from the CPs is negative
For any value of $\overline{H} \in (-1, 0]$, the IR constraint make the strategies of the players coupled and the game can have continuum of equilibria as it is the case with general coupled constrained games \cite{GNEP}. However, with continuum of equilibria we will be faced with equilibrium selection problem and a systematic comparison of CP utilities under the two regime is not possible. We thus set $\overline{H}=-1$ under which the IC constraint ensures that the IR constraint always holds and hence the objective of the CPs are no more jointly constrained. As we will see in the subsequent sections, this avoids the continuum of equilibrium. We note that even after relaxing the IR constraint the problem is still challenging to analyze,  but makes it possible to compare equilibrium utilities of all players under both the regimes. 


We say that a contract profile $(\beta_i,i\in \mathcal{N})$ is an equilibrium if no CP has an incentive for unilateral deviation from its contract. In the following we superscript the quantities computed at equilibrium with NN and N when they are associated with non-neutral regime and neutral regime respectively.


 

%% file: MultiSymmetry.tex
In this section we consider the symmetric case where revenue per unit
demand for all the CPs is the same, i.e., $r_1=r_2, \ldots,=r_n := r.$
In other words, the CPs are symmetric with regards to the ability to
monetize their content.
In this setting, we analyze the equilibrium contracts arising in the
neutral as well as non-neutral regime, and the resulting surplus of
the CPs and the ISP. Our results highlight, surprisingly, that even
when the CPs are symmetric, the imposition of neutrality actually
shrinks the surplus of all parties involved. Moreover, this `loss of
surplus' becomes more pronounced as the number of CPs $n$ grows.


\subsection{Non-neutral regime}
In the non-neutral regime, it is easy to see that when
$\overline{H}=-1$, the interactions between each CP and the ISP are
decoupled. The optimization problem in (\ref{eqn:OptNonNeutral}) for
$CP_i, i\in N,$ after substituting the optimal effort simplifies to:
\begin{align*}
	&\max_{\beta_i\in [0,1]} \hspace{2mm} (1-\beta_i)r\log\left(\max\left(\frac{\beta_ir}{c},1\right)\right).
\end{align*}
Moreover, we note that it is only interesting to consider the case
$r > c.$ Indeed, since the monetization resulting from ISP effort
$a_i$ for CP~$i$ equals $r\log(1+a_i),$ the marginal monetization is
at most $r.$ Thus, if $r \leq c,$ it is not worthwhile for CPs to make
any investments to grow the demand.


The following result characterizes the equilibrium contracts between
each CP and the ISP. The contracts are expressed in terms of the
LambertW function computed on its principle branch, denoted as
$W(\cdot)$ (see~\cite{corless1996lambertw}).
\begin{thm}
  \label{thm:equillibrium-NN}
  If $r > c,$ the equilibrium contract between $\CPi, i \in \mathcal{N}$ and the
  ISP is given by 
  \begin{equation} \label{eqn:SymmetricBetaNN}
    \beta^{NN}:=\beta^{NN}_i= \frac{1}{W\left(\frac{r}{c}e\right)}.    
  \end{equation}
\end{thm}
Since $W(\cdot)$ in strictly increasing and $W(e) = 1,$ it follows
that $\beta^{NN} \in (0,1)$ when $r/c > 1.$ Moreover, note that
equilibrium fraction $\beta^{NN}$ of CP revenue that is shared with the
ISP is a strictly decreasing function of the ratio $r/c,$ as might be
expected.

Using Theorem~\ref{thm:equillibrium-NN}, one can easily characterize
the equilibrium effort of the ISP as well as the surplus of each
agent.
\begin{cor}
  \label{cor:equillibrium-NN}
  Assume $r > c.$ The equilibrium effort for each $\CPi, i \in \mathcal{N}$ put
  by the ISP is given by
  \begin{equation} \label{eqn:SymmetricEffortNN}
    a^{NN}:=a^{NN}_i= \frac{r\beta^{NN} }{c}-1 > 0.
  \end{equation}
  The equilibrium surplus of $\CPi, i \in \mathcal{N}$ is given by
  \begin{equation}
    \label{eqn:SymmetricCPUtilityNN}
    \UCPi^{NN}=(1-\beta^{NN}) r\log(a^{NN}+1)
    =\frac{(1-\beta^{NN})^2}{\beta^{NN}}r > 0.
  \end{equation}
  Finally, the equilibrium surplus of the ISP is given by
  \begin{equation}
    \label{eqn:SymmetricISPUtilityNN}
    \UISP^{NN}=nr+nc-2n\beta^{NN}r > 0.
  \end{equation}
\end{cor}

Note that so long as $r>c,$ the equilibrium contracts award each CP
and the ISP a positive surplus.


\subsection{Neutral regime}

We now consider the neutral regime. The CPs are still assumed to be
symmetric, only the ISP is now \emph{constrained} to make the same
investment decision for all CPs, i.e., $a_1=\ldots=a_n:=a $.  The
surplus of $\CPi$ in this case, after substituting the
optimal ISP effort simplifies to:
\begin{align*}
  (1-\beta_i)r\log\left(\max\left(\frac{\sum_{j=1}^{n}\beta_jr}{nc},1\right)\right).
\end{align*} 
Since the surplus of each CP in the neutral regime depends on the
actions of all CPs, we seek contract profiles $(\beta^N_i, i \in N)$ that
constitute a \emph{Nash equilibrium} between CPs. These equilibria are
characterized completely in the following theorem. As before, the only
scenario of interest is $r >c.$

\begin{thm}
  \label{thm:equillibrium-N}
  Consider the neutral regime with $r > c.$ In this case, only
  symmetric Nash equilibria exist. When $1 < r/c \leq n,$ there are
  two Nash equilibria, $\boldsymbol{0},$ and $(\beta^N_i, i \in \mathcal{N}),$
  where
  \begin{equation}
    \label{eqn:SymmetricBetaN}
    \beta_i^N = \beta^N := \frac{1}{nW\left(\frac{r}{nc}e^{1/n}\right)}.
  \end{equation}
  When $r/c > n,$ $(\beta^N_i, i \in \mathcal{N})$ is the only Nash equilibrium,
  where $\beta_i^N$ is given by \eqref{eqn:SymmetricBetaN}.
\end{thm}

Note that when $1 < r/c \leq n,$ unlike in the non-neutral regime,
making no contributions to the ISP, resulting in zero suplus for all
parties, is an equilibrium between the CPs. The other equilibrium,
given by \eqref{eqn:SymmetricBetaN}, results in a positive surplus for
all parties (as is shown in the following corollary). In the remainder
of this section, we will refer to this latter equilibrium as the
\emph{non-zero} equilibrium.

\begin{cor}
  Consider the neutral regime, with $r > c.$ Under the non-zero
  equilibrium:
  \begin{itemize}
  \item The effort for $\CPi, i \in N$, put by the ISP is given by
    \begin{equation} \label{eqn:SymmetricEffortN}
      a^N:=a_i^N(n)=\max\left(\frac{\beta^N r}{c}-1,0\right).
    \end{equation}
  \item The surplus of $\CPi, i\in N$ is given by
    \begin{equation}
      \label{eqn:SymmetricCPUtility}
      \UCPi^{N}=(1-\beta^{N}) r\log(a^{N}+1)
      =\frac{\left(1-\beta^{N}\right)^2}{n\beta^{N}}r > 0.
    \end{equation}
  \item The surplus of ISP is given by
    \begin{equation}
      \label{eqn:SymmetricISPUtility}
      \UISP^{N}=r+nc-(n+1)\beta^{N}r > 0.
    \end{equation}
  \end{itemize}
\end{cor}


\subsection{Neutral regime v/s Non-neutral regime}

Having now characterized the equilibrium contracts and the surplus of
each CP and the ISP under the neutral and the non-neutral regime, we
are now in a position to compare the two. As the following result
shows, the non-neutral regime is actually better \emph{for all
  parties} as compared to the neutral regime.
\begin{thm}
  \label{thm:ComparisonSymmetric}
  Suppose $r > c,$ and $n \geq 2.$ In the symmetric case, at
  equilibrium, the following statements hold.
  \begin{enumerate}
  \item CPs share a higher fraction of their revenue with the ISP in
    the non-neutral regime, i.e., $\beta^{NN} > {\beta^{N}}.$
  \item The effort by the ISP for each CP is higher in the non-neutral
    regime, i.e., $a^{NN} > {a^{N}}$
  \item The surplus of each CP is higher in the non-neutral regime,
    i.e., $\UCPi^{NN}> \UCPi^N$ for all $i \in N$
  \item The surplus of the ISP is higher in the non-neutral regime,
    i.e., $\UISP^{NN} > \UISP^N$.
  \end{enumerate}
\end{thm}
The above result highlights that, surprisingly, constraining the ISP
to be neutral is actually sub-optimal for all parties, even when the
CPs are symmetric. In other words, the non-neutral regime is actually
preferable to the ISP as well as the CPs. Intuitively, the reason for
this \emph{tragedy of the commons} is that the imposition of
neutrality skews the payoff landscape for each CP, such that the
`benefit' of any additional investment it makes gets `shared' across
all CPs. This induces the CPs to commit smaller fractions of their
revenues to the ISP, which in turn results in a lower ISP effort, and
a lower demand growth for all CPs. Indeed, as we show below, this
effect gets further magnified with an increase in the number of CPs.

%
%
%
%
%
%

\subsection{The effect of number of CPs}

In the non-neutral regime, the interactions between the different CPs
and the ISP are decoupled, implying that the impact of scaling $n$ is
trivial. Thus, we now study the impact of scaling $n$ in the
neutral regime, on the equilibrium ISP effort, and the surplus of each
agent. Note that when $n=1,$ the neutral and the non-neutral regime
coincide. Our main result is the following.


\begin{thm}
  \label{thm:EffectOfCPsize}
  Suppose that $r > c.$ In the neutral regime, the non-zero
  equilibrium satisfies the following properties.
\begin{enumerate}
\item $\beta^N$ is a strictly decreasing function of $n.$
\item The effort by the ISP for each CP ($a^N$) is a strictly
  decreasing function of $n,$ even though the total effort ($na^N$) by
  the ISP is a strictly increasing function of $n.$  
\item The surplus of each CP is a strictly decreasing function of $n,$
  and $\lim_{n \rightarrow \infty} U^N_{CP_i}(n) = 0.$
\item The surplus of the ISP is eventually strictly decreasing in $n,$
  and $\lim_{n \rightarrow \infty} U^N_{ISP}(n) = 0.$
\end{enumerate} 
\end{thm}

\begin{figure}[h!] 
	\centering
	{\includegraphics[scale=0.225]{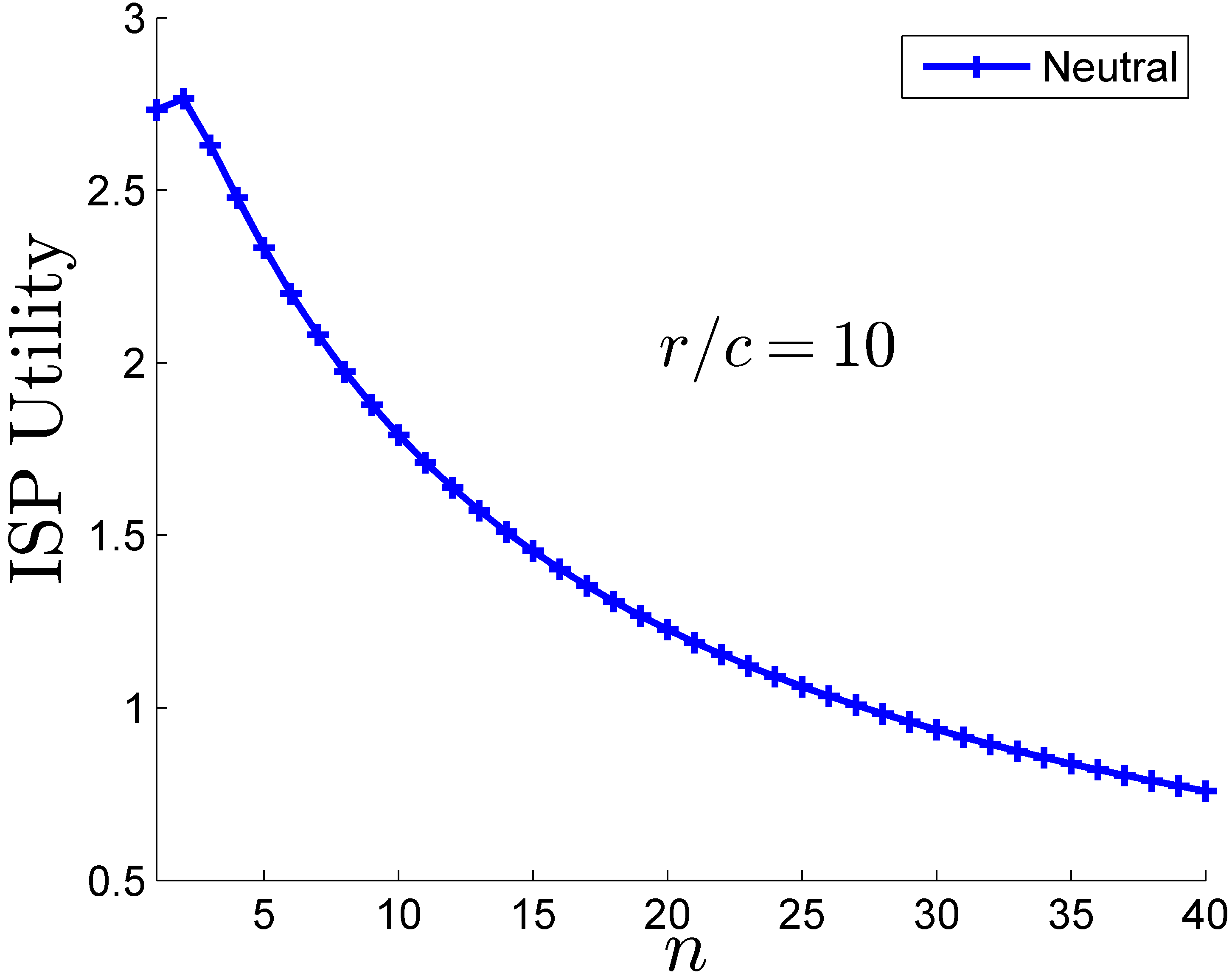}%
	}
	{\includegraphics[scale=0.225]{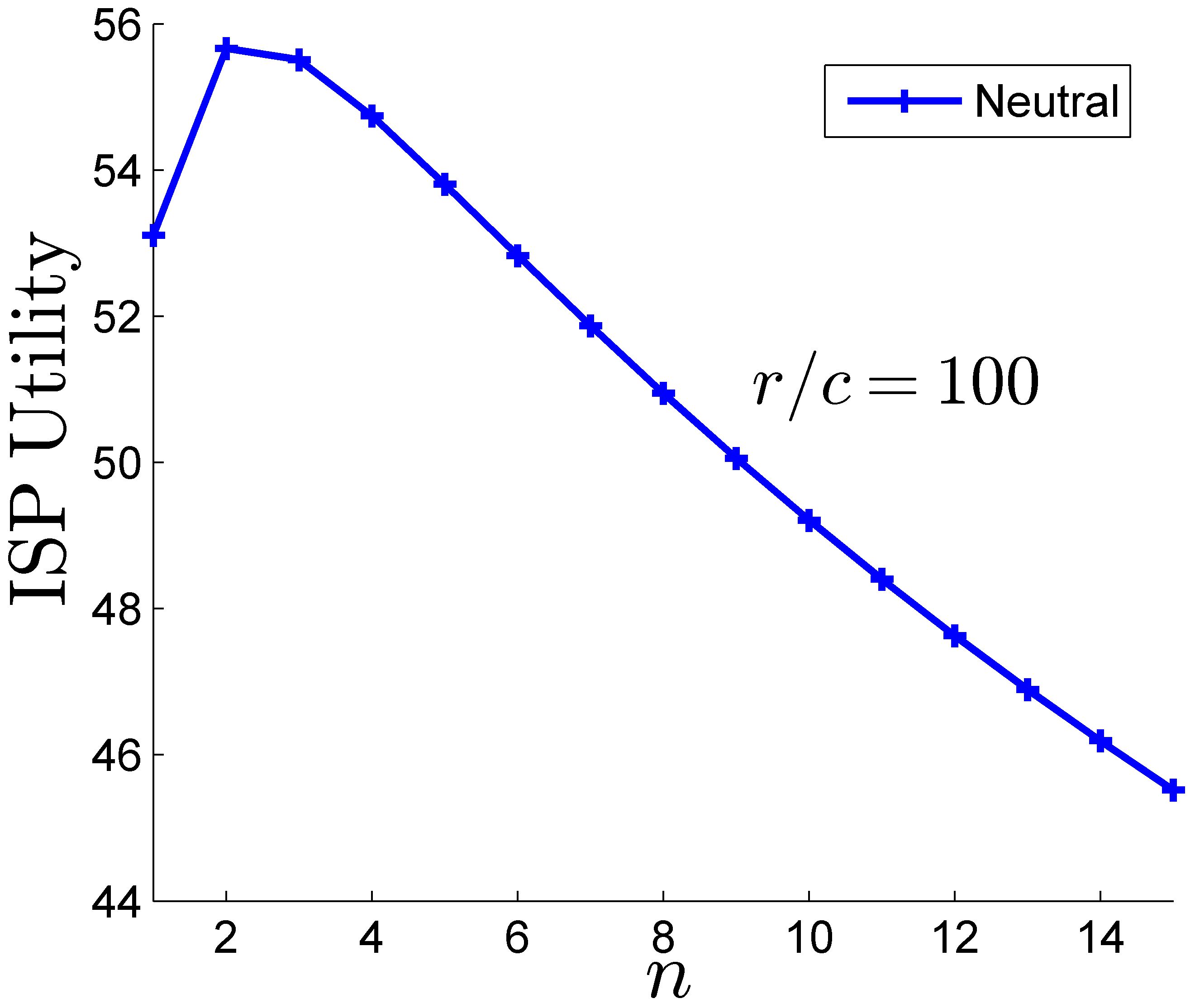}%
	}
	{\includegraphics[scale=0.225]{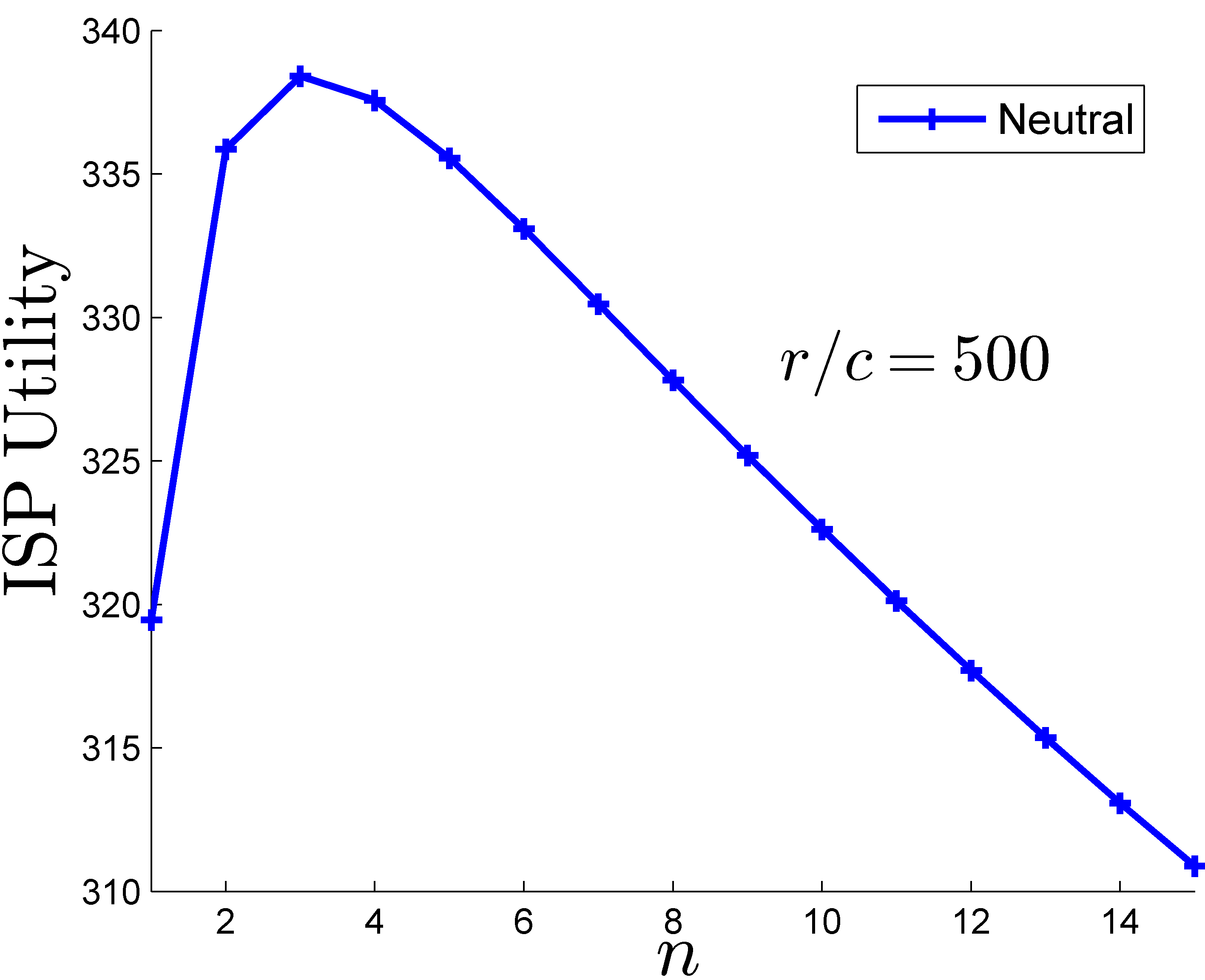}%
	}
	\caption {ISP utility in the neutral regime
          as $n$ varies for different $r/c$.}
        \label{fig:neutral_n_variation_ISP}
\end{figure}

Theorem~\ref{thm:EffectOfCPsize} highlights that an increase in the
number of CPs further exacerbates the sub-optimality of the neutral
regime for the CPs as well as the ISP. As before, the explanation for
this is that with increasing $n,$ the surplus resulting from an
additional contribution by any $CP$ gets `split' further, thus
disincentivising the CPs from offering a significant fraction of their
revenues to the ISP. The variation of ISP utility as a function of $n$ is depicted in Figure \ref{fig:neutral_n_variation_ISP} for different values of $r/c$. In all cases, the utility first increase for some $n$ and the decreases thereafter.

%% file: AsymmetricCase.tex
In this section we study the asymmetric case where monetizing power of all the CPs need not be the same, i.e., $r_i\neq r_j$ for $i\neq j$.
Our interest in this section is to understand how disparity in the
monetizing power influences preference of the players for the neutral
and non-neutral regimes. We focus on the case with two
CPs ($n=2$) and without loss of generality assume that monetizing
power of $\mbox{CP}_1$ is more than that of $\mbox{CP}_2,$ i.e., $r_1>r_2.$ We refer to $\mbox{CP}_1$ as dominant and  $\mbox{CP}_2$ as non-dominant.

Recall that the objective of the $CP_i, i\in \mathcal{N}$ in the
non-neutral regime can be expressed as
\begin{equation*}
\max_{\beta_i \in [0,1]} \hspace{2mm} (1-\beta_i)r_i\log\left(\max\left(\frac{\beta_ir_i}{c},1\right)\right),
\end{equation*}
and in the neutral regime it can be expressed as
\begin{equation*}
  \max_{\beta_i \in [0,1]} \hspace{2mm} (1-\beta_i)r_i\log\left(\max\left(\frac{\sum_i\beta_ir_i}{nc},1\right)\right).
\end{equation*} 
As discussed in Section \ref{sec:MultipleSymmetric}, the case
$r_i/c\leq 1$ for all $i \in \mathcal{N}$ is not interesting as none
of the CPs would have the incentive to contribute towards ISP effort.
Thus, in this section, we restrict ourselves to the case where
$r_i/c>1$ for at least one $i \in \mathcal{N}$.  The following results
characterize the equilibrium contracts for the neutral and the
non-neutral regime.\footnote{The characterization of equilibrium
  contracts can actually be done for any $n;$ see Appendix~\ref{app:AsymmetricMultipleCPs}.}

\subsection{Equilibrium contracts}

In the non-neutral regime, the interactions between each CP and the
ISP remain decoupled, and thus the equilibrium contracts follow easily
from Theorem~\ref{thm:equillibrium-NN}.
\begin{cor}
  \label{thm:AsymBetaNN}
  In the non-neutral regime, the equilibrium contract
  $(\beta_1^{NN},\beta_2^{NN})$ is as follows:
  $$\beta_i^{NN} = \begin{cases}
      0  &\mbox{  if }  \frac{r_i}{c} \leq 1, 
      \\
      \frac{1}{W\left(\frac{r_i}{c}e\right)}  &\mbox{  if }  \frac{r_i}{c} > 1.
    \end{cases}$$
\end{cor}
\noindent Note that when $\frac{r_i}{c} \leq 1,$ the equilibrium
contract between $\CPi$ and the ISP is not uniquely defined, since any
$\beta_i \in [0,1]$ would result in zero surplus to $\CPi.$

Next, we characterize equilibrium contract in the neutral regime.
\begin{thm} \label{thm:AsymBetaN} Consider the neutral regime, with
  $r_1 > r_2.$ If $r_1/c \leq 2$ then
  $(\beta_1^N,\beta_2^N)=(0,0)$. If $r_1/c > 2,$ then the equilibrium
  contract is given by:\\ \begin{equation}
    (\beta_1^N,\beta_2^N)=\begin{cases} (\overline{\beta}_1,
      \overline{\beta}_2) &\mbox{ if } \frac{r_1+r_2}{r_1-r_2} >
      2W\left(\frac{r_1+r_2}{4c}\sqrt{e}\right), \\ \left(\frac{1}{
          W\left(\frac{r_1}{2c}e\right)},0\right) &\mbox{
        otherwise}, \end{cases} \label{eqn:NeutralConds} \end{equation}
  where \begin{equation}\nonumber
    \overline{\beta}_1=\frac{r_1+r_2}{4r_1W\left(\frac{r_1+r_2}{4c}\sqrt{e}\right)}-\frac{r_2-r_1}{2r_1},
    \overline{\beta}_2=\frac{r_1+r_2}{4r_2
      W\left(\frac{r_1+r_2}{4c}\sqrt{e}\right)}-\frac{r_1-r_2}{2r_2}.  \end{equation}
\end{thm}

When $r_1/c \leq 2,$ the equilibrium contract is not unique, though
the outcome is that ISP effort equals zero. When $r_1/c > 2,$ the
equilibrium contract is unique, and at least one CP (specifically,
$\mbox{CP}_1$) is guaranteed to contribute a constant fraction of her
revenue to the ISP. Note that when $r_1/c \in (1,2],$ there is no CP
contribution in the neutral regime, even though there is in the
non-neutral regime. 

To interpret the equilibrium when $r_1/c > 2,$ let $r_1^*:=r_1^*(r_2)$
denote the value of $r_1$ that satisfies the following relation for a
given $r_2$
$$\frac{r_1+r_2}{r_1-r_2} = 2W\left(\frac{r_1+r_2}{4c}\sqrt{e}\right).$$
For $r_1\leq r_1^*$ the condition in (\ref{eqn:NeutralConds}) holds where revenue shared by both the CPs is strictly positive,
i.e., $\beta_i^{N}>0$ for all $i\in \N$. For $r_1 > r_1^*$ the condition in (\ref{eqn:NeutralConds}) fails
in which only the $\mbox{CP}_1$'s share is strictly
positive and $\mbox{CP}_2$ does not share anything, i.e.,
$\beta_1^N>0$ and $\beta_2^N=0$. Further, it is easy to verify that
$r_1^*$ is monotonically increasing in $r_2$ and $r_1^*> r_2$.

\subsection{Comparison between Neutral and Non-neutral regimes}

Having characterized the equilibrium contracts in both regimes, we
compare and contrast the neutral and non-neutral regimes in the
remainder of this section. We begin by comparing the equilibrium
contracts, followed by CP/ISP utility, social utility, and finally ISP
effort.

\subsubsection{\textbf{Contracts}}
The following proposition provides a comparison of the equilibrium
contracts in both the regimes.

\begin{figure*}[!t]
	\centering
	\subfloat[]{\includegraphics[scale=0.05]{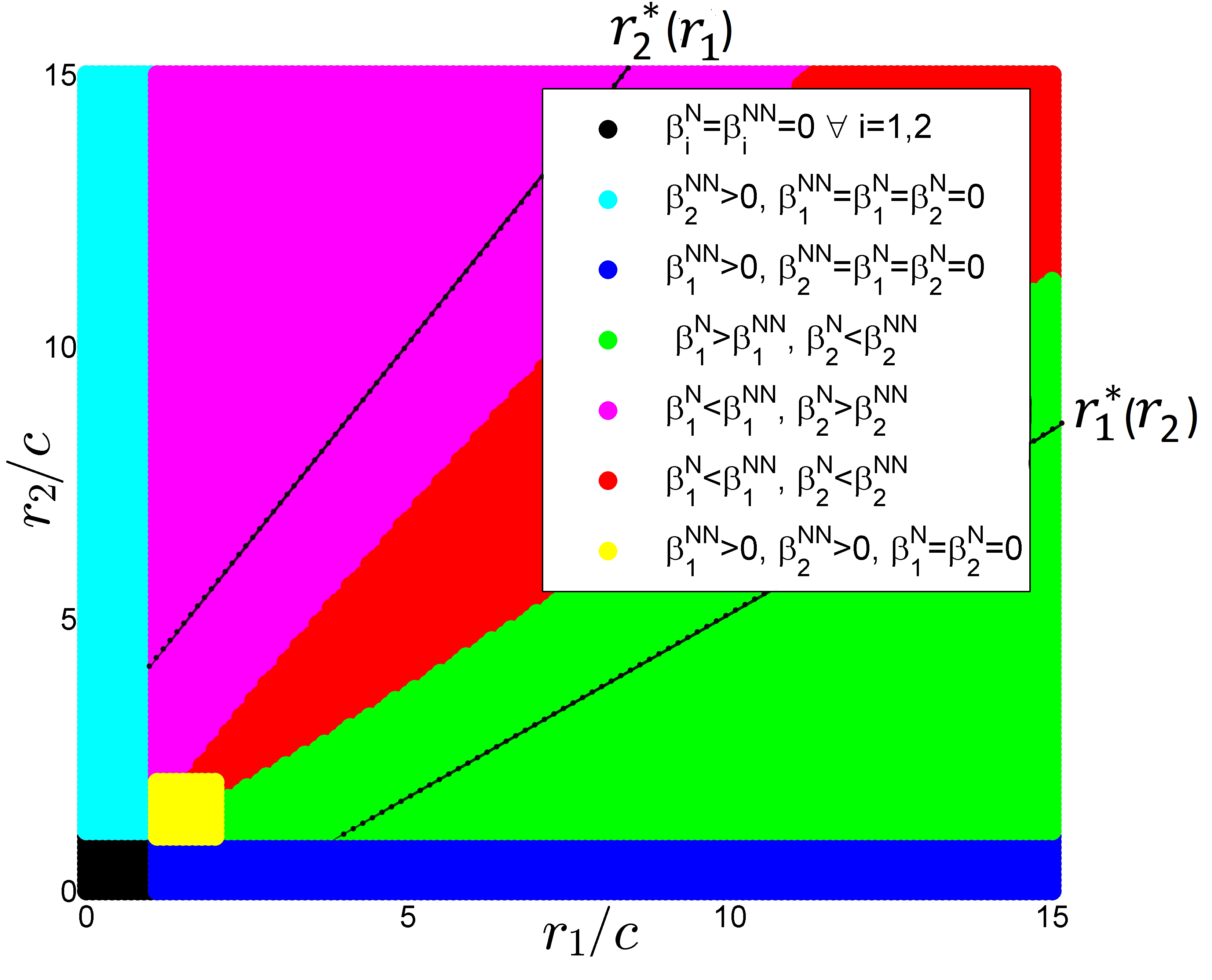}\label{fig:betascatter}}
	\hspace{0.4cm}
	\subfloat[]{\includegraphics[scale=0.3]{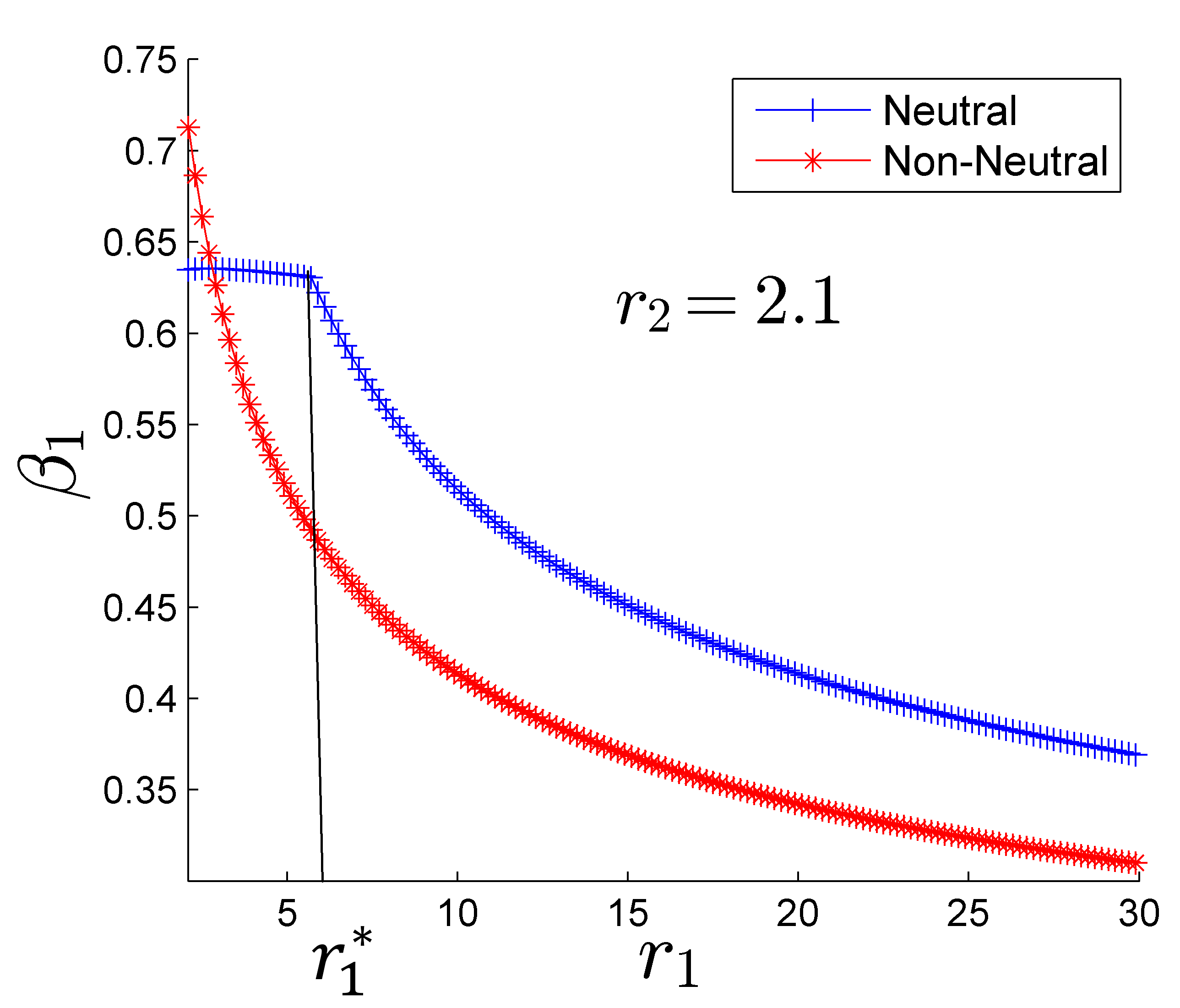}
		\label{fig:beta1a}}
	\hspace{0.4cm}
	\subfloat[]{\includegraphics[scale=0.3]{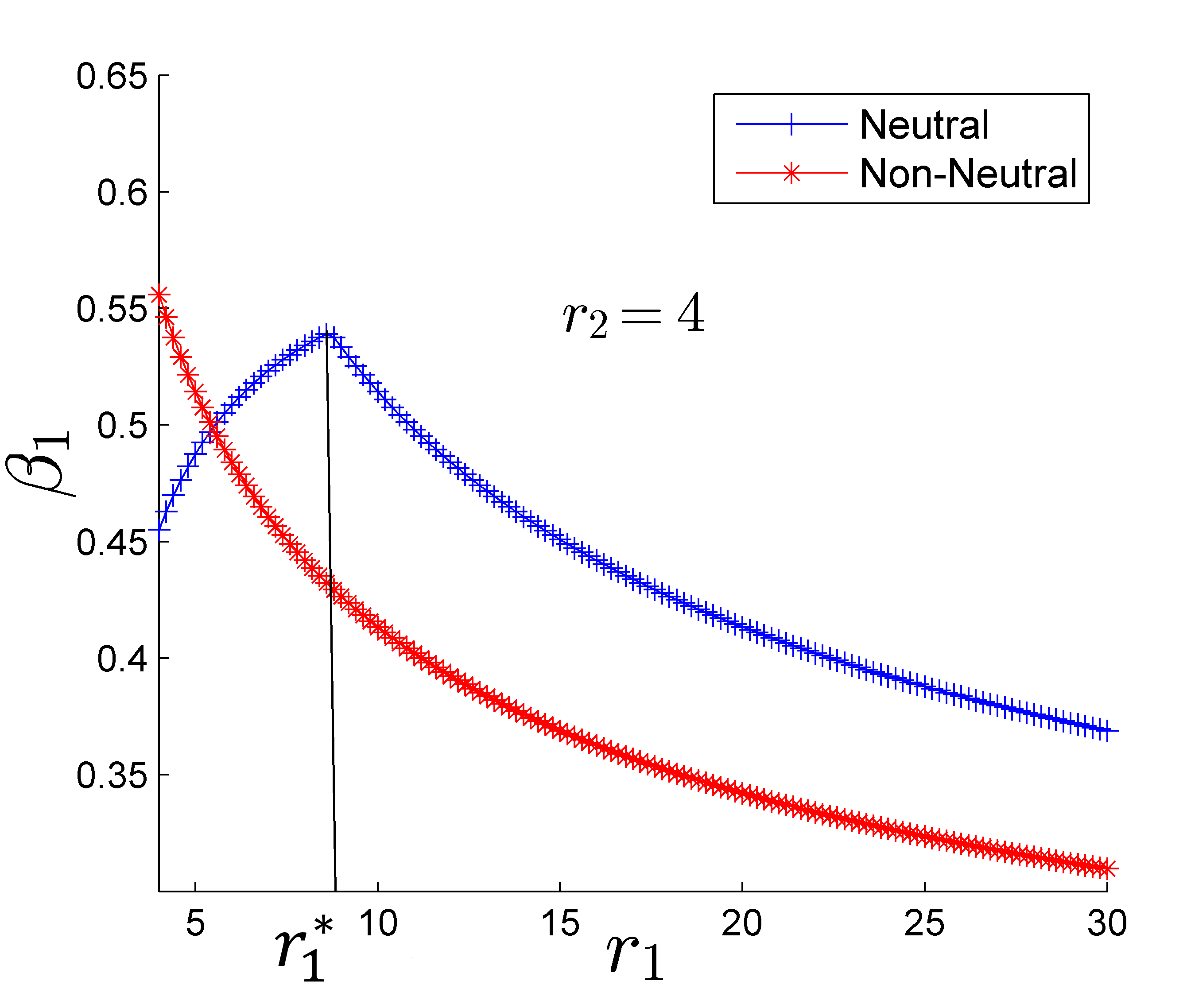}
		\label{fig:beta1e}}
	\caption {Fig.~\ref{fig:betascatter} gives scatter-plot of $\beta$s. Figs.~\ref{fig:beta1a} and \ref{fig:beta1e} shows variation of equilibrium $\beta_1$ vs $r_1$ under neutral and non-neutral regime.}
	\label{fig:beta}
\end{figure*}

\begin{figure*}[!h]
	\centering
	
	\subfloat[]{\includegraphics[scale=0.3]{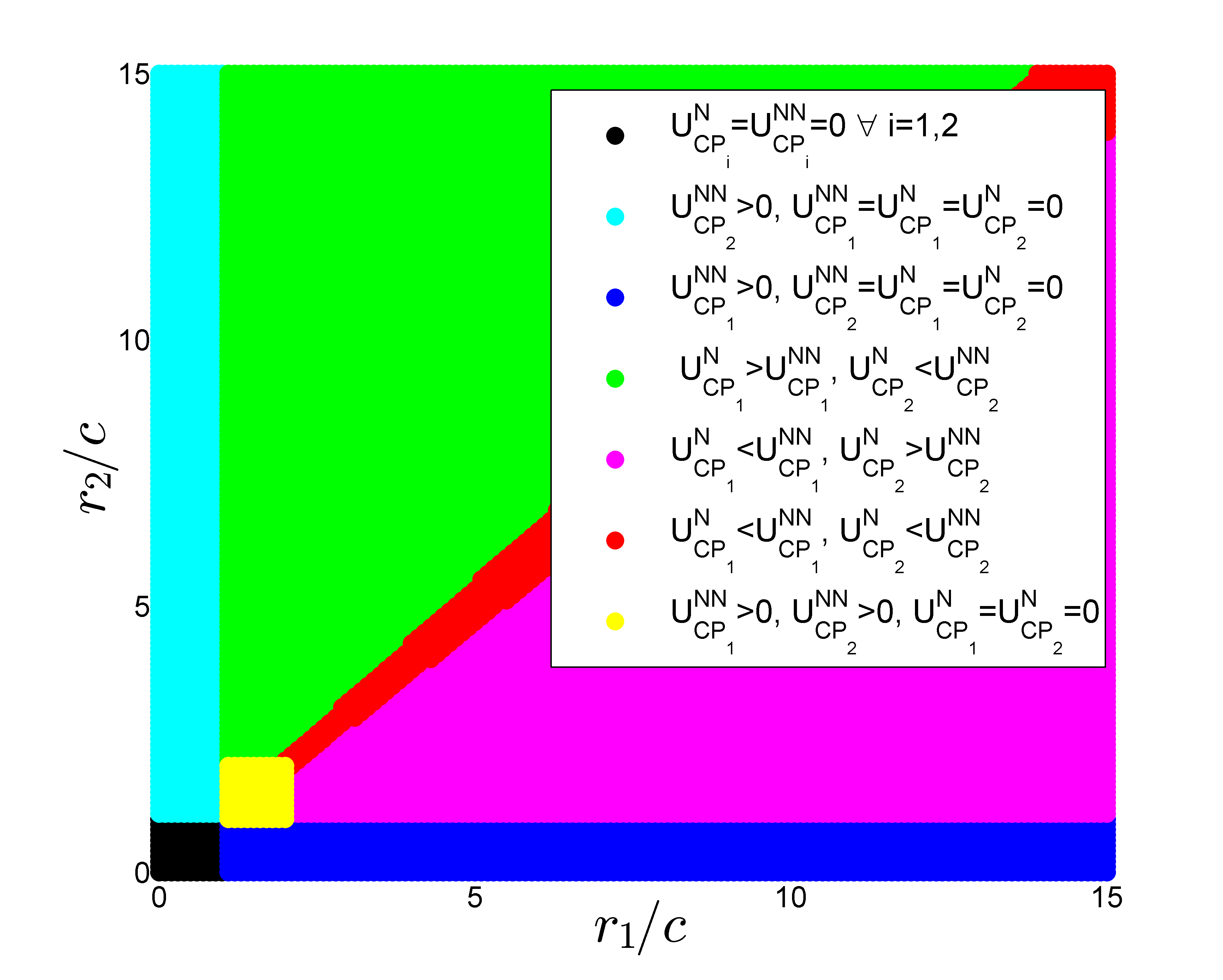}		\label{fig:CPutilityscatter}}
	\hspace{0.4cm}
	\subfloat[]{\includegraphics[scale=0.3]{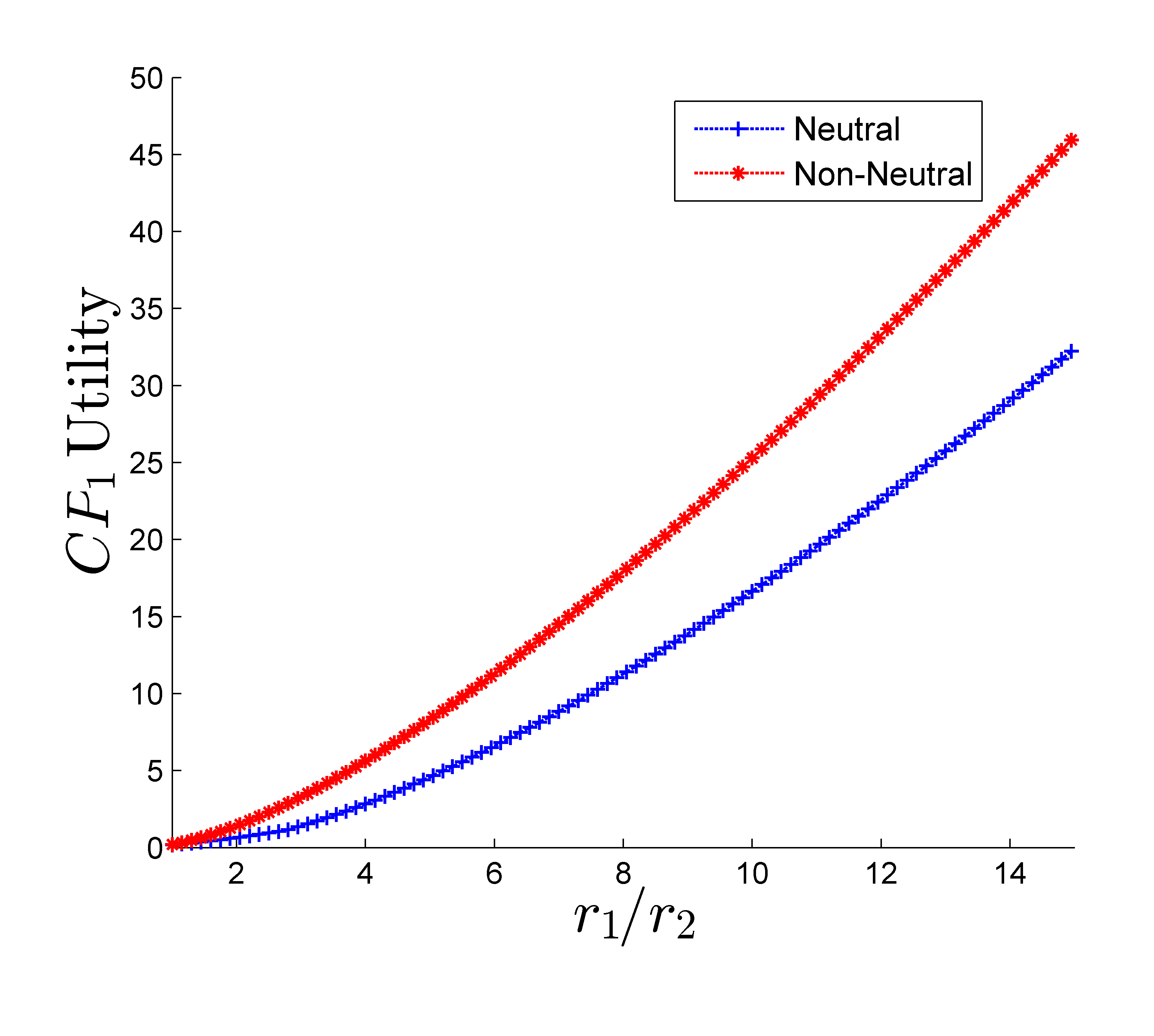}\label{fig:CP1utility}}
	\hspace{0.4cm}
	\subfloat[]{\includegraphics[scale=0.3]{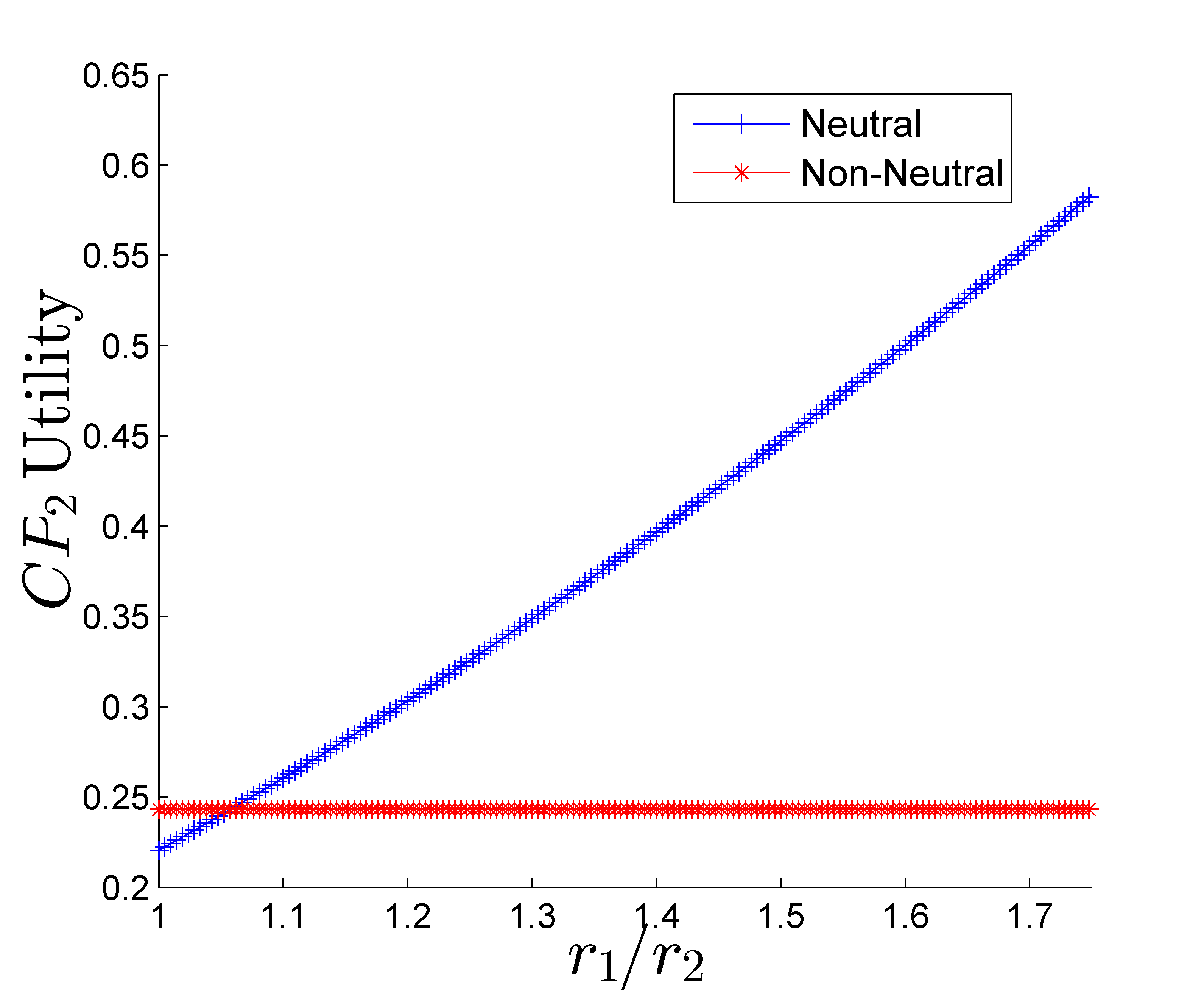}\label{fig:CP2utility}}
	\caption {Fig.~\ref{fig:CPutilityscatter} shows scatterplot for the CP utilities at equilibrium. Figs.~(\ref{fig:CP1utility}) \& (\ref{fig:CP2utility}) compare $\mbox{CP}_1$ utility in both regimes as $r_1$ varies.}
	\label{fig:CPUtility}
\end{figure*}

\begin{proposition} 
  \label{prop:contracts}
  Fix $r_2 > 0.$ We have
  \begin{itemize}
  \item For $r_1> r_2$, $\beta_2^{NN} \geq \beta_2^{N}$.
    Moreover, $\beta_2^{N}$ decreases in $r_1$.
  \item For $r_1\geq r_1^*$,
    $\beta_1^{N}>\beta_1^{NN}$. Moreover, $\beta_1^{N}$
    decreases in $r_1$ for $r_1 \geq r_1^*$.
  \end{itemize}	
\end{proposition}
The conclusions of Proposition~\ref{prop:contracts} are summarised in
the scatter plot in~Fig.~\ref{fig:betascatter}. Note that the
non-dominant CP always contributes a smaller fraction of its revenue
in the neutral regime. With the dominant CP, the contribution factor
is larger in the neutral regime when the revenue rates are highly
asymmetric (see the green region in Figure~\ref{fig:betascatter}, and
larger in the non-neutral regime when the revenue rates are symmetric
(see the red region in Figure~\ref{fig:betascatter}). A sufficient
condition for the former is $r_1\geq r_1^*(r_2).$ The latter
observation is of course consistent with Theorem \ref{thm:ComparisonSymmetric}, which dealt
with the case of perfect symmetry.

Proposition~\ref{prop:contracts} also establishes monotonicity
properties of the sharing contracts of $\mbox{CP}_1$  in the neutral regime in $r_1$ for a fixed $r_2.$ While
$\beta_2^{N}$ descreasess in $r_1,$ $\beta_1^{N}$
eventually decreasing in $r_1;$ see Figs.~\ref{fig:beta1a}
and~\ref{fig:beta1e}. Note that $\beta_1^{N}$ can actually be
increasing with respect to $r_1$ when the revenue rates are nearly
symmetric, in contrast with the non-neutral setting.




\subsubsection{\textbf{Utility of CPs}}
The following proposition characterizes preference of the CPs
for the neutral and non-neutral regime.
%

\begin{proposition}
	\label{prop:CPUtility}
	Fix an $r_2$. We have
	\begin{itemize}
		\item For all $r_1>r_2$,  $\mbox{CP}_1$ prefers the non-neutral regime.
		\item For all $r_1\geq r_1^*$,  $\mbox{CP}_2$ prefers the neutral regime.
	\end{itemize}
\end{proposition}
Figure~\ref{fig:CPutilityscatter} shows the scatter plot utilities of
the players in both the regimes. Note that the dominant CP has higher
utility in the non-neutral regime as can be observed from the red and
magenta regions. This is because in the neutral regime, the dominant
CP is `forced' to pay for capacity investments that also benefit the
non-dominant~CP. Indeed, note that in the region $r_1\geq r_1^*$, the
dominant CP shares a smaller fraction of its revenue in the
non-neutral regime, but still ends-up with a higher
utility. Interestingly, the non-dominant CP obtains a higher utility
in the neutral regime when the asymmetric revenue rates are too separated (see the
pink region in Fig.~\ref{fig:CPutilityscatter}) A sufficient condition
for this is $r_1\geq r_1^*.$ This is of course due to the
`subsidization' it receives from the dominant CP. On the other hand,
when the revenue rates are nearly symmetric, even the non-dominant CP
prefers the non-neutral regime, once again consistent with
Theorem~\ref{thm:ComparisonSymmetric}. The above observations further illustrated in
Figs.~\ref{fig:CP1utility} and~\ref{fig:CP2utility}.



\subsubsection{\textbf{ISP utility}}
\begin{figure}[h!]
	\centering
	{\includegraphics[scale=0.29]{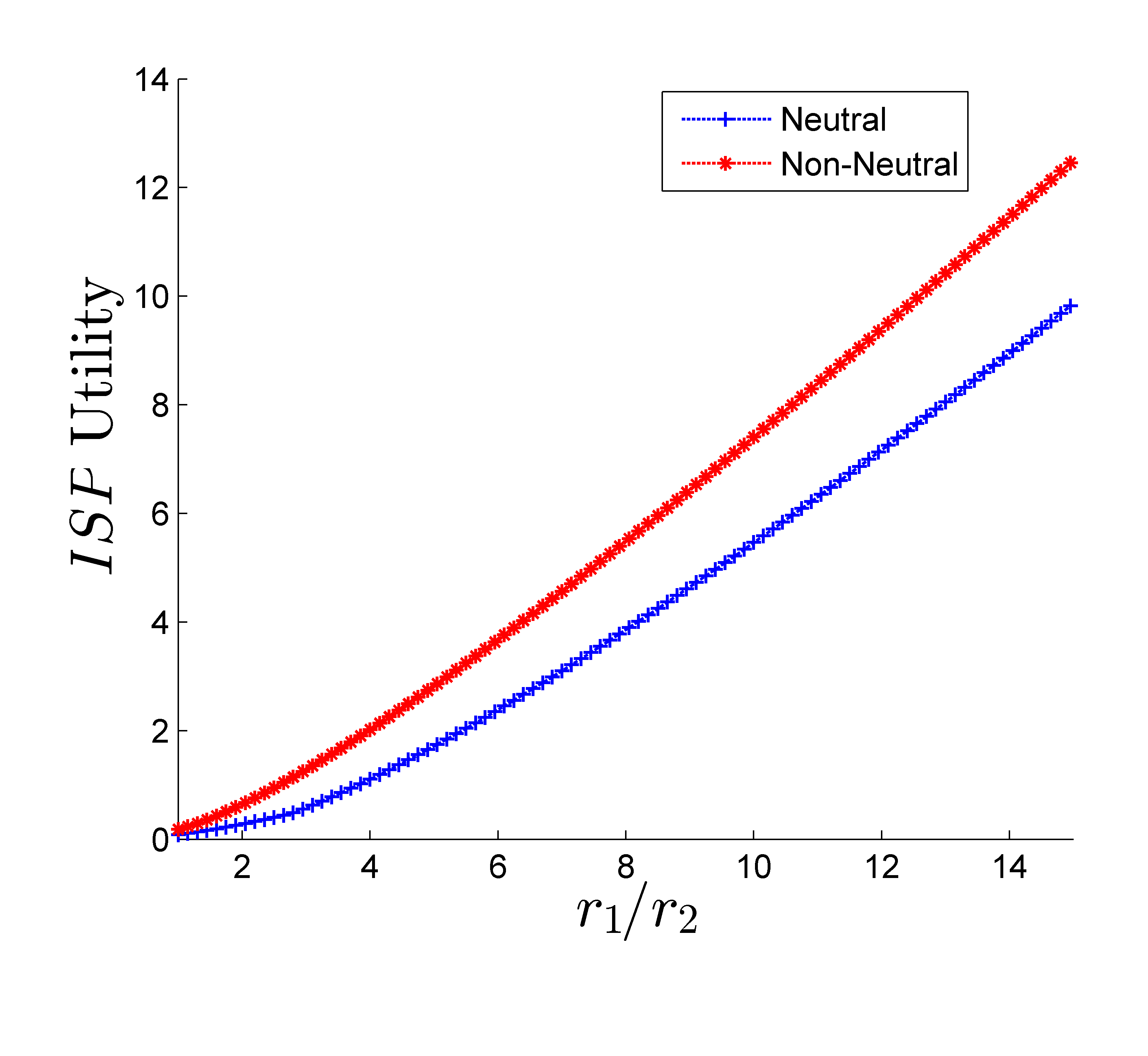}%
	}
	{\includegraphics[scale=0.29]{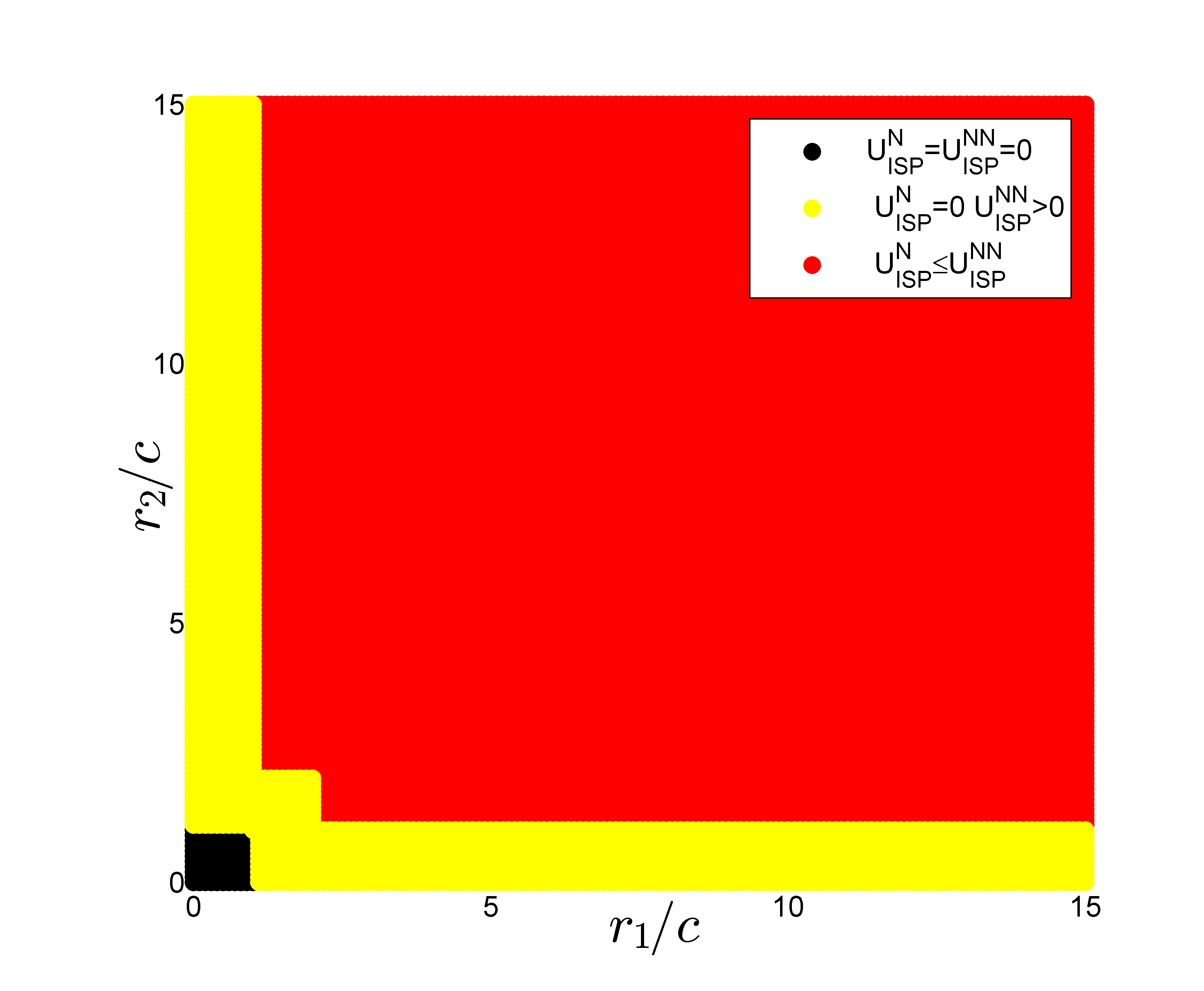}%
	}
	\caption {The first figure compares ISP utility in neutral and non-neutral with $c=1 r_2=2$. The second figure gives a scatter plot. }
	\label{fig:ISPUtility}
\end{figure}

We next compare utility of the ISP in the non-neutral and neutral
regime. For simplicity, we take ISP utility to be the expected revenue
given as $\UISP=\mathbb{E}[\sum_{i} s(X_i)-ca_i]$ (ignoring the
risk-sensitive utility defined before). Its value in the non-neutral
regime is given by:
\[\UISP^{NN}=(1-2\beta_1^{NN})r_1+(1-2\beta_2^{NN})r_2+2c,\]
and in the neutral regime for all $r_1\geq r_1^*$ is given by:
\[\UISP^{N}=(1-2\beta_1^{N})r_1+2c.\] 
The utility for  $r_1< r_1^*$ in the neutral regime is cumbersome and we skip its expression. The following lemma demonstrates the ISPs earnings are higher in the non-neutral regime when monetization power of the dominant CP is much larger than the other, i.e., $r_1$ is much larger than $r_2$.
\begin{lemma} \label{lma:ISPutilitycomp}
	There exist $r_1^b>r_1^*$, such that for all $r_1>r_1^b$ the ISP's utility is higher in the
	 non-neutral regime.
\end{lemma}
A general comparison of ISP utility in the two regimes is not analytically
tractable. We give a numerical illustration in Figure  \ref{fig:ISPUtility}. As seen in the first figure, utility of ISP in the non-neutral
regime is higher than in the neutral regime for all $r_1$ for a  given $r_2$ and $c$.  
Scatter plot in the second figure shows that this observation extends over the entire parameter range.
\subsubsection{\textbf{Social Utility}}
The social utility in the non-neutral and neutral regimes are given, respectively, as follows:
\begin{align*}
&SU^{NN}=U_{CP_1}^{NN}+U_{CP_1}^{NN}+U_{ISP}^{NN}\\
=&r_1\log\left(\frac{\beta_1^{NN}r_1}{c}\right)+r_2\log\left(\frac{\beta_2^{NN}r_2}{c}\right)-(\beta_1^{NN}r_1+\beta_2^{NN}r_2)+2c
\end{align*}
and
\begin{align*}
&SU^N=U_{CP_1}^N+U_{CP_1}^N+U_{ISP}^N\\
&=(r_1+r_2)\log\left(\frac{\beta_1^Nr_1+\beta_2^Nr_2}{c}\right)-(\beta_1^Nr_1+\beta_2^Nr_2)+2c.
\end{align*} 
\begin{figure}[h!] 
	\centering
	{\includegraphics[scale=0.29]{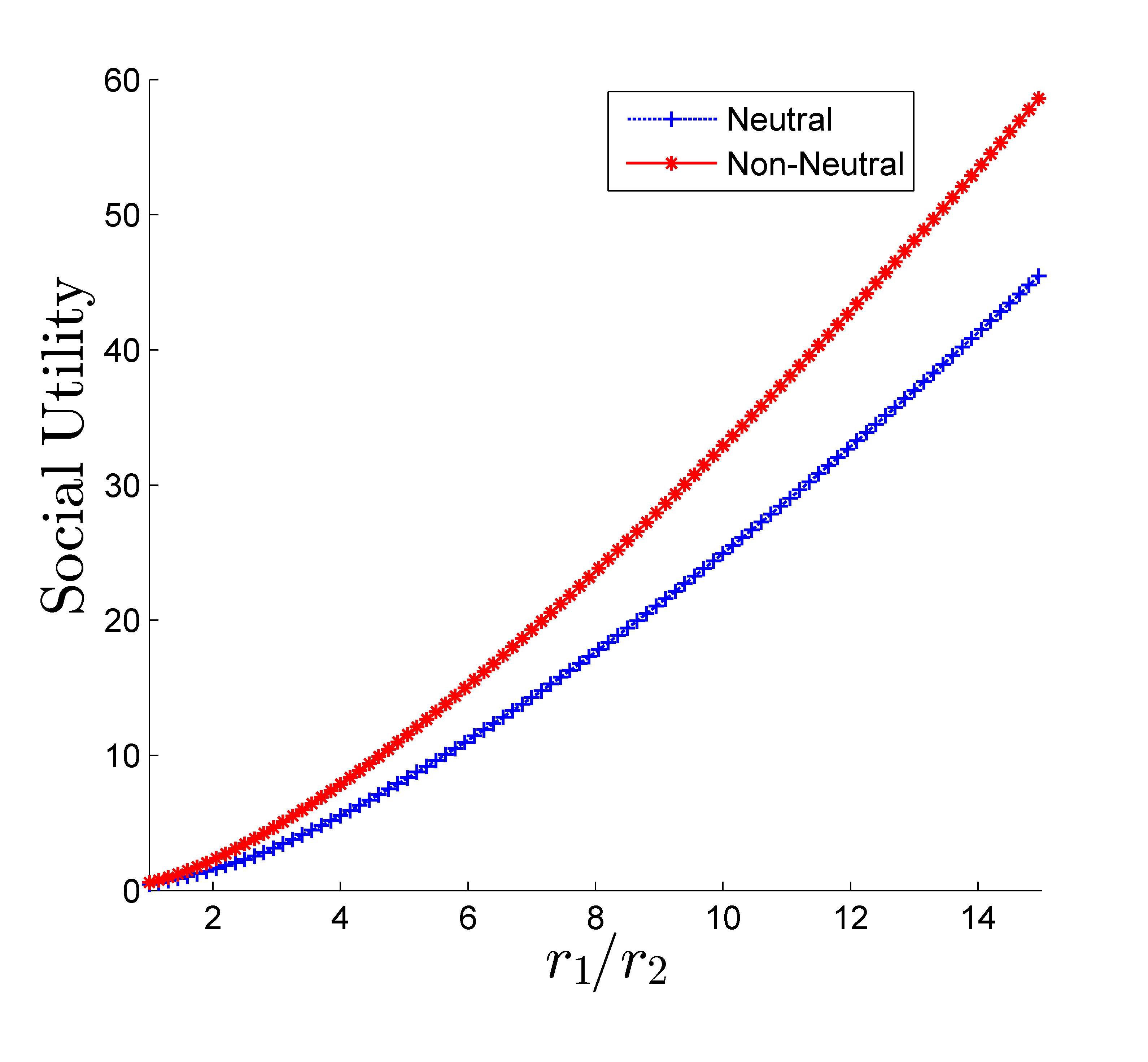}%
	}
	{\includegraphics[scale=0.29]{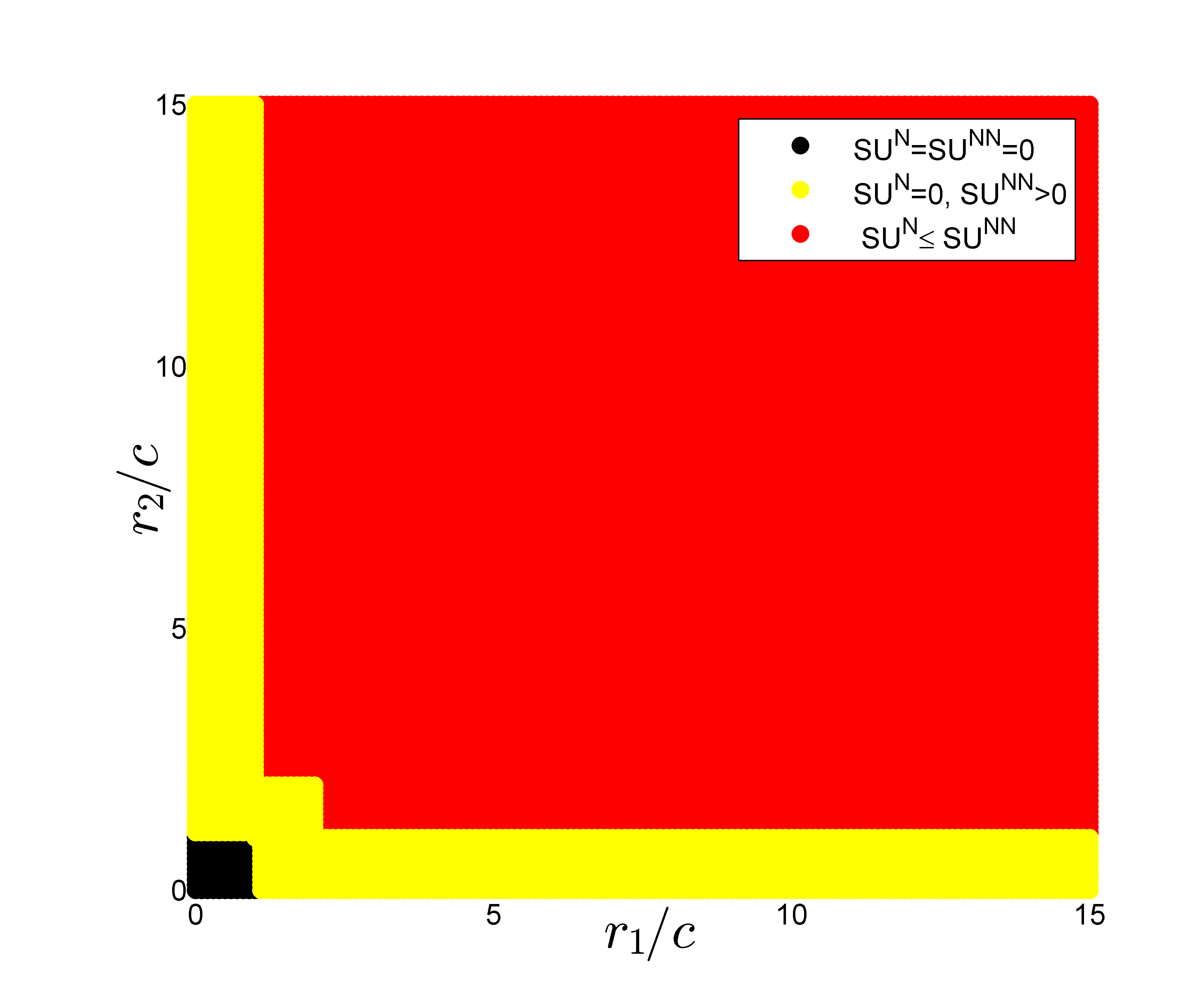}%
	}
	\caption {Comparison of social utility between neutral and non-neutral regime for $c=1, r_2=2$ and scatter plot. }
	\label{fig:SocialUtility}
\end{figure}
As it is not easy to compare the social utilities analytically, we resort to numerical comparison of the utilities in Figure  \ref{fig:SocialUtility}. As seen social utility in the non-neutral regime dominates that in the neutral regime for all values of $r_1$ for a given $r_2$ and $c$. The scatter plot in the second part of the figure shows that the observation continue to hold for all parameters.
\subsubsection{\textbf{Total Effort by ISP}}
Finally we compare the total effort by ISP for CPs in the non-neutral and neutral regime given, respectively, as follows
\begin{align*}
A^{NN}&=a_1^{NN}+a_2^{NN}=\frac{\beta_1^{NN}r_1}{c}-1+\frac{\beta_2^{NN}r_2}{c}-1\\
&=\frac{\frac{r_1}{c}}{W(\frac{r_1}{c}e)}+\frac{\frac{r_2}{c}}{W(\frac{r_2}{c}e)}-2 \;\; \forall \; r_1>r_1^*,
\end{align*}
and
\begin{align*}
A^{N}=2a^{N}=2\left(\frac{\beta_1^{N}r_1}{2c}-1\right)=\frac{\frac{r_1}{c}}{W(\frac{r_1}{2c}e)}-2 \hspace{.75mm}\forall \hspace{.75mm}r_1>r_1^*
\end{align*}
\begin{figure}[!t] \label{fig:TotalEffort}
	\centering
	{\includegraphics[scale=0.29]{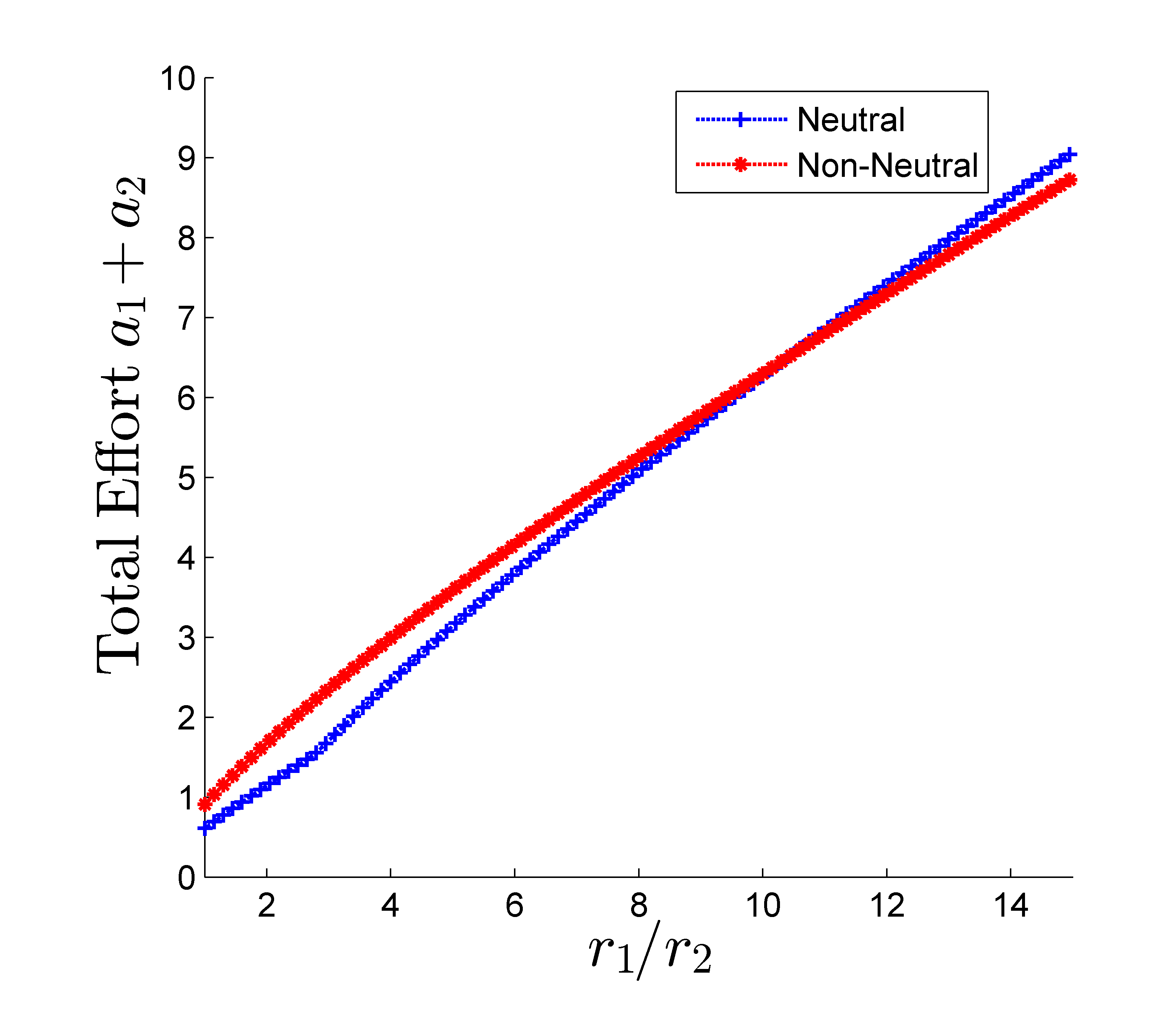}%
	}
	{\includegraphics[scale=0.35]{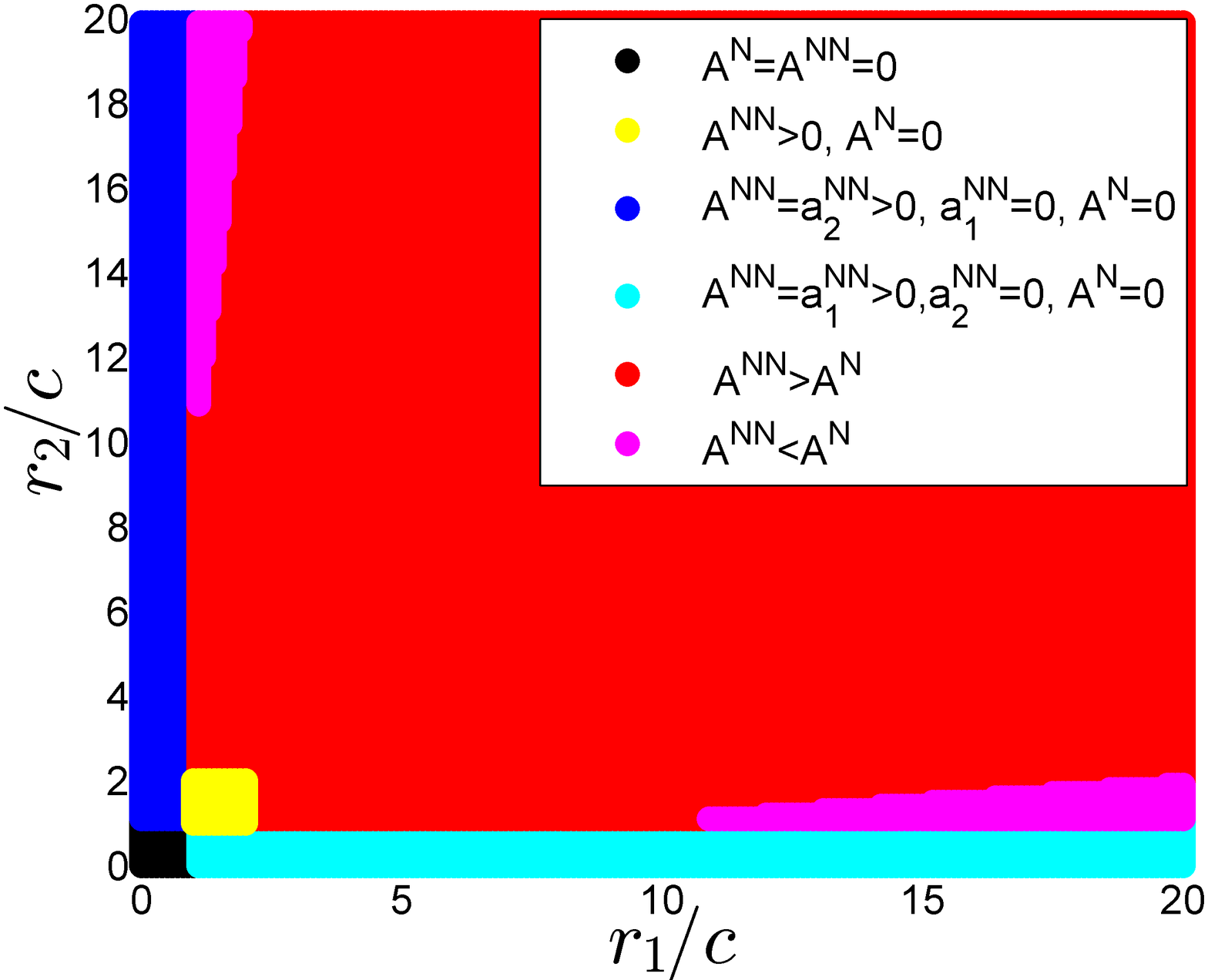}%
	}
	\caption{Scatter plot for comparison between total effort (investment) by ISP in non-neutral and neutral regime over different range of $r_i/c$}
\end{figure}
\begin{lemma} \label{lma:totaleffortcomp}
	There exist a threshold $r_1^a>r_1^*$ such that total effort by ISP is higher in neutral regime than in the  non-neutral. The threshold satisfies the following:
\end{lemma}
The threshold $r_1^a$ is given by following equation:
$$r_1^a \left(\frac{1}{W(\frac{r_1^a}{2c}e)}-\frac{1}{W(\frac{r_1^a}{c}e)}\right)=\frac{{r_2}}{W(\frac{r_2}{c}e)}$$
It can be seen from above equation that $r_1^a$ is monotonically increasing in $r_2$.\\

The above lemma implies that total effort by ISP in the neutral regime becomes higher when there is high asymmetry between CP's revenue per click rates. 

%% file: Appendix.tex
\subsection{Proof of Theorem \ref{thm:equillibrium-NN}}
\label{app:equilibrium-NN}
From CP$_i$ optimization problem, it can be observed that for $r/c < 1$ $\UCPi=0$ for all $i$. Henc
e no CP has an incentive to share a fraction of their revenue with the ISP and $\beta_i=0 \; \forall i\in N$  is the equilibirum. Now assume $r/c \geq 1$. For this case the optimial value of $\beta_i$ will be such that $r\beta_i/c\geq 1$ and the optimization problem of $\CPi$ reduces to
\[
\max_{\beta_i\in [0,1]} \hspace{2mm} (1-\beta_i)r\log\left(\frac{\beta_ir}{c}\right).
\]
The first order optimality condition $\partial U_{CP_i}/\partial \beta_i =0$ then gives:
$$\log\left(\frac{\beta_ir}{c}\right)=\frac{1-\beta_i}{\beta_i} \hspace{1.5mm} \forall i=1,2,...,n.$$
Solving the first order conditions for each CP$_i$, we get:
\begin{align*}
&	\frac{1-\beta_i}{\beta_i}=\log\left(\frac{\beta_i r }{c}\right)	\implies \frac{1}{\beta_i}=\log\left(\frac{\beta_i re }{c}\right)\\
&	\implies e^\frac{1}{\beta_i}=\frac{\beta_i re }{c} 	\implies \frac{1}{\beta_i}e^\frac{1}{\beta_i}=\frac{r e}{c}
\end{align*}
Using the definition of the LamebertW funtion we get
 \[\frac{1}{\beta_i}=W\left(\frac{r }{c}e\right) \implies \beta_i=\frac{1}{W\left(\frac{r }{c}e\right)}\]
 Hence we get equilibrium contract given in (\ref{eqn:SymmetricBetaNN}). \hfil\IEEEQED

\subsection{Proof of Theorem \ref{thm:equillibrium-N}}
\label{app:equillibrium-N}
Recall the objective of $\CPi, i\in N$
\begin{align*}
&\max_{\beta_i \in [0,1]} \hspace{2mm} (1-\beta_i)r\log\left(\max\left(\frac{\sum_{j=1}^{n}\beta_jr}{nc},1\right)\right).
\end{align*} 
First assume that $r/c<1$. In this case for any given $(\beta_1,\beta_2,\ldots, \beta_{i-1}, \beta_{i+1}, \ldots, \beta_n )$, best response of $\CPi$ is to set $\beta_i=0$. Thus $\beta_i=0 \; \forall i \in N$ is an equilibrium.

\noindent
Next consider the case 
$1 \leq  r/c < n$. Fix an $i\in N$ and assume $\beta_j=0$ for all $j\neq i$. Then the object of $\CPi$ simplifies to
\begin{align*}
&\max_{\beta_i \in [0,1]} \hspace{2mm} (1-\beta_i)r\log\left(\max\left(\frac{\beta_ir}{nc},1\right)\right),
\end{align*} 
and the best response of $\CPi$ is to set $\beta_i=0$. Hence $\beta_i=0$ for all $i \in N$ is an equilibrium. We next look for a non-zero equilibrium. By symmetry, it must be such that $\beta_1=\beta_2\ldots=\beta_n \in (0,1]$. Further, at equilibrium it must be the case that $\sum_{i=1}^n \beta_jr/nc\geq 1$, otherwise CPs have incentive to deviate to make their share zero. Writing the first order condition for the optimaization problem of  $\CPi, i \in N$, i.e., 
\begin{align*}
&\max_{\beta_i \in [0,1]} \hspace{2mm} (1-\beta_i)r\log\left(\frac{\sum_j \beta_jr}{nc}\right),
\end{align*} 
we get $$\log\left(\frac{\sum_{j=1}^n\beta_jr}{nc}\right)=\frac{1-\beta_i}{\sum_{j=1}^n\beta_j}.$$
Setting $\beta_1=\beta_2, \ldots,=\beta_n=\beta$ we have
$$\log\left(\frac{\beta r}{c}\right)=\frac{1-\beta}{n\beta}.$$
Simplyfying the above as earlier in the format of LambertW function we get $\beta=\frac{1}{nW(\frac{r}{nc}e^{1/n})}$. \hfil\IEEEQED

\noindent
For the case $r/c \geq n$, $\beta_i=0, \forall i \in N$ at equilibrium is not arise, however the equilibrium $\beta=\frac{1}{nW(\frac{r}{nc}e^{1/n})}$ still holds. This completes the proof.
\subsection{Proof of Theorem \ref{thm:ComparisonSymmetric}}
\label{app:ComparisonSymmetric}
\noindent
{\it Part 1:} When $r/c\leq  1, \beta^{NN}=\beta^N=0$ and the relation $\beta^{NN} \geq \beta^N$ holds trivially. In the range $1 < r/c \leq  n$, two equilibria are possible in the neutral regime, $\beta^{N}=0$  or $\frac{1}{nW(r/ce^{1/n})}$. If $\beta^{N}=0$ is the equilibrium, again the relation holds trivially. Consider the case when $\beta^{N}=\frac{1}{nW(r/ce^{1/n})}$ is the equilibrium for $1 < r/c $. Define $b:=r/c$ and $f(b)=\frac{\beta^{NN}}{\beta^N}$.
\begin{align*}
\lim _{b\rightarrow 1}f(b)&=\lim _{b\rightarrow 1}\frac{nW({\frac{b}{n}}e^{\frac{1}{n}})}{W(e)}\\
&=\frac{nW({\frac{1}{n}}e^{\frac{1}{n}})}{W(e)} =\frac{n.\frac{1}{n}}{1} = 1\hspace{1mm} (\text{using  } x=W(xe^x))
\end{align*}
The limit holds as the equlibrium definition holds for all $b>1$ and $W$ is continuous at $b=1$.
Also, $f(b)$ is monotonically increasing in $b$ forall $b>1$ as
$$\frac{\partial f(b)}{\partial b}=\frac{nW(be^{\frac{1}{n}}/n)}{b W(be)}\left[\frac{W(be)-W(be^{\frac{1}{n}}/n)}{(1+W(be))(1+W(be^{\frac{1}{n}}/n))}\right] >0 \hspace{.25mm} \forall b>1
$$
Hence $\beta^{NN}> \beta^{N}$.  It holds similarly for the case $r/c>n$.

\noindent
{\it Part 2:}  Since investment decision by ISP is monotonically increasing in the share in gets fromt the CPs (from Eqns. (\ref{eqn:SymmetricEffortNN}) and (\ref{eqn:SymmetricEffortN}), by Part $1$ it is clear that ISP  make more investment in non-neutral regime as compared to neutral regime.

\noindent
{\it Part 3:}  In both non-neutral and neutral regime equilibrium effort, $a$ for given $\beta$ is $a+1=\max\left(\frac{\beta r}{c},0\right)$.

\begin{figure}
	\centering
	\includegraphics[scale=.13]{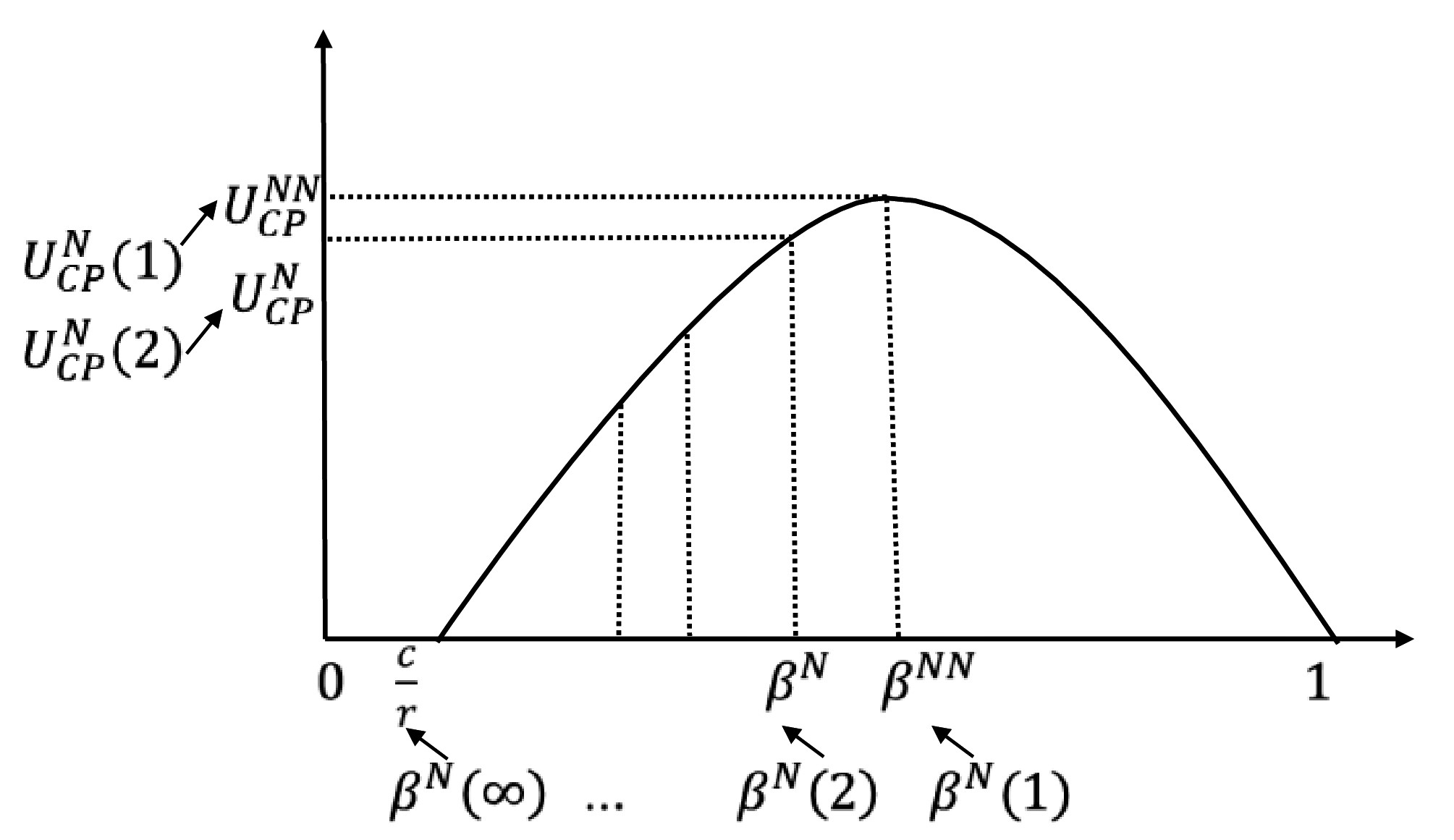}
	\caption{Utility of CP vs $\beta$}
	 \label{fig:SymmetricCPvsBeta}
\end{figure}
Now, in both non-neutral and neutral regimes, each CP's utility at equilibrium is the same function given by $(1-\beta) r \log \left(\max\left(\frac{\beta r}{c},0\right)\right)$
which is concave in $\beta \in (c/r,1)$ and from Part 1 we have that $\beta^{N} \leq \beta^{NN}$. This implies that $U_{CP}^{N} \leq U_{CP}^{NN}$ (seen Fig. \ref{fig:SymmetricCPvsBeta})

\noindent
{\it Part 4:}  $$U^{NN}_{ISP}=[n\beta^{NN} r \
\log(a^{NN}+1)-nc(a^{NN})]$$
Substituting the value of $a^{NN}(\beta^{NN})=\frac{\beta^{NN}r}{c}-1$ in the second term of above expression, we get:
\begin{equation}\label{UISPnn}
U_{ISP}^{NN}=\{n\beta^{NN} r\left[\
\log(a^{NN}+1)-1\right]+nc\}
\end{equation} 
Similarly,
\begin{equation} \label{UISPn}
U_{ISP}^{N}=[n\beta^{N} r\left[\log(a^{N}+1)-1\right]+nc]
\end{equation}
Now, from Part $1$ and $2$, we know that $\beta^{NN} \geq {\beta^{N}}$ and $a^{NN} \geq {a^{N}}$, respectively. Thus, by comparing Eqns. (\ref{UISPnn}) and (\ref{UISPn}), we have that $U_{ISP}^{NN}\geq U_{ISP}^{N}$. \hfil\IEEEQED

\subsection{Proof of Theorem \ref{thm:EffectOfCPsize}}
\label{app:EffectOfCPsize}
{\it Part 1:} Considering $n$ to be continuous variable,
\begin{align*}
	\frac{\partial \beta^N(n)}{\partial n}&=\frac{1}{n^2W\left(\frac{r}{nc}e^{1/n}\right)}\left[-1+\frac{1+\frac{1}{n}}{1+W\left(\frac{r}{nc}e^{1/n}\right)}\right]
\end{align*}
Now, $\beta^N(n)$ decreases with $n$ \textit{iff} $	\frac{\partial \beta^N}{\partial n}<0$
\begin{align*}
\iff\left[-1+\frac{1+\frac{1}{n}}{1+W\left(\frac{r}{nc}e^{1/n}\right)}\right]<0\iff& \left[\frac{1+\frac{1}{n}}{1+W\left(\frac{r}{nc}e^{1/n}\right)}\right] <1\\
	\iff {1+\frac{1}{n}}<1+W\left(\frac{r}{nc}e^{1/n}\right)\iff  &1<nW\left(\frac{r}{nc}e^{1/n}\right)
\end{align*}
it holds as $\beta^N(n)=\frac{1}{nW\left(\frac{r}{nc}e^{1/n}\right)}<1 \implies nW\left(\frac{r}{nc}e^{1/n}\right)>1$.

\noindent
{\it Part 2:} Effort of ISP for each CP is decreasing following directly as  $\beta^N$ is decreasing in $n$. The total effort of ISP is $A^N(n)=na^N(n)=n\left(\frac{\beta^N r}{c}-1\right)$. In the following we show that $A^N(n+1)>A^N(n)$ for any $n$. We have
\begin{equation}
A^N(n+1)>A^N(n)\iff (n+1)\beta^N(n+1)-n\beta^N(n)>\frac{c}{r}
\end{equation}We prove that the above inequality holds in two part.\\
{\it Part (i):} We first prove that $g(n)=n\beta^N(n)$ is concave in $n$, which implies the difference $ (n+1)\beta^N(n+1)-n\beta^N(n)$ shrinks as $n$ increases.
Now, 
$$g(n)=n\beta^N (n)=\frac{1}{W\left(\frac{b}{n}e^\frac{1}{n}\right)}; \mbox{ where } b=\frac{r}{c}$$
It is clear that $g(n)$ is increasing in $n$. Treating $n$ as continuous variable, we have
$$\frac{\partial f(n)}{\partial n}=\frac{n+1}{n^2W\left(\frac{b}{n}e^\frac{1}{n}\right)\left(1+W\left(\frac{b}{n}e^\frac{1}{n}\right)\right)}>0$$
$$\frac{\partial^2 f(n)}{\partial n^2}=\frac{1}{n^4W\left(\frac{b}{n}e^\frac{1}{n}\right)\left(1+W\left(\frac{b}{n}e^\frac{1}{n}\right)\right)^2}\times$$
$$\left \{(n+1)^2\frac{\left(1+2W\left(\frac{b}{n}e^\frac{1}{n}\right)\right)}{\left(1+W\left(\frac{b}{n}e^\frac{1}{n}\right)\right)}-n(n+2)\left(1+W\left(\frac{b}{n}e^\frac{1}{n}\right)\right) \right\}$$
Now, $f(n)$ is strictly concave in $n$ \textit{iff} $\frac{\partial^2 f(n)}{\partial n^2}<0$ 
$$\iff\frac{(n+1)^2}{n(n+2)}<\frac{\left(1+W\left(\frac{b}{n}e^\frac{1}{n}\right)\right)^2}{\left(1+2W\left(\frac{b}{n}e^\frac{1}{n}\right)\right)}$$
After cross multiplying and expanding, we get
$$\frac{1+2W\left(\frac{b}{n}e^\frac{1}{n}\right)}{nW\left(\frac{b}{n}e^\frac{1}{n}\right)+2W\left(\frac{b}{n}e^\frac{1}{n}\right)}<nW\left(\frac{b}{n}e^\frac{1}{n}\right)$$
We know that $1/\beta^N=nW\left(\frac{a}{n}e^\frac{1}{n}\right)>1$ at equilibrium,therefore LHS$<$1 and RHS$>$1. Thus, the above inequality holds . \\
{\it Part (ii):} Now, we show that $(n+1)\beta^N(n+1)-n\beta^N(n) \rightarrow c/r \hspace{2mm} \text{as} \hspace{2mm}n\rightarrow \infty $.
Consider asymptotic expansion of LambertW function, $W(x)=x-x^2+o(x^2)=x(1-x+o(x))$
$$\frac{1}{W(x)}=\frac{1}{x(1-x+o(x))}=\frac{1}{x}\cdot\left(1+x+o(x)\right)=\frac{1}{x}+1+o(1)$$
second equality comes by using $\frac{1}{1-x}=1+x+o(x)$.Now, 
\begin{align} \nonumber
	f(n)&=n\beta^N(n)=\frac{1}{W\left(\frac{r}{nc}e^\frac{1}{n}\right)}=\frac{nc}{re^\frac{1}{n}}+1+o(1)\\
	&=\frac{nc}{r}\left(1-\frac{1}{n}+o\left(\frac{1}{n}\right)\right)+1+o(1)=\frac{nc}{r}-\frac{c}{r}+1+o(1)
\end{align}
third equality comes by using $e^{-x}=1-x+o(x)$.Therefore,
$$f(n+1)-f(n)=\frac{(n+1)c}{r}-\frac{nc}{r} \rightarrow \frac{c}{r}\hspace{2mm} \text{as}\hspace{2mm} n\rightarrow \infty$$

\noindent
{\it Part 3:} For any $n$ each CP's utility for the given $a$ at equilibrium, is given by $(1-\beta) r \log \left(\max\left(\frac{\beta r}{c},0\right)\right)$.\\
We know that the above function is concave in $\beta \in (c/r,1]$ and from Part $1$, as $n$ increases $\beta^N(n)$ decreases and approaches $c/r$ when $n\rightarrow \infty$ $ \left(\frac{1}{nW\left(\frac{r}{nc}e^\frac{1}{n}\right)}=\frac{1}{n\left(\frac{r}{nce^{1/n}}-\left(\frac{r}{nce^{1/n}}\right)^2-o\left(\frac{1}{n^2}\right)\right)}\rightarrow \frac{c}{r}, \hspace{.4mm}\text{as} \hspace{.4mm} n\rightarrow \infty\right)$. This implies that $U_{CP}^{N}(n)$ decreases as $n$ increases (see Fig. \ref{fig:SymmetricCPvsBeta}). Also it can be seen that as $n\rightarrow \infty, U^N_{CP}\rightarrow 0$.

\noindent
{\it Part 4:}
 \begin{align*}
 	U_{ISP}^{N}&=n(\beta^{N}r\log({a^{N}}+1)-ca^N)\\&=n\left(\beta^{N}r\log\left(\frac{\beta^{N}r}{c}\right)-c\left(\frac{\beta^N r}{c}-1\right)\right)\\
 	&=r+nc-(n+1)\beta^Nr
 \end{align*}
Utility of ISP from each CP is
 \begin{equation}
\frac{ U_{ISP}^{N}}{n}=\beta^{N}r\log({a^{N}}+1)-ca^N
 \end{equation}
 Substituting the value of $a^N=\frac{\beta^Nr}{c}-1$ in the second term, we get $	\frac{ U_{ISP}^{N}}{n}=\beta^{N}r \left(\log({a^{N}}+1)-1\right)+c$.
%
Since $\beta^N$ ans $a^N$ decreases with increase in $n$, it is clear from above expression that utility of ISP from each CP also decreases with increase in number of CPs.\\
 
Now, $U_{ISP}(n)^N$ increases with increase in $n$ \textit{iff}
 \begin{align*}
 	\frac{U_{ISP}(n)^N}{\partial n}&>0 \hspace{2mm} \text{(considering $n$ to be continuous)}\\
 	c-\frac{1}{n^2W\left(\frac{r}{nc}e^\frac{1}{n}\right)}&\left(\frac{n^2+n+1-nW\left(\frac{r}{nc}e^\frac{1}{n}\right)}{n\left(1+W\left(\frac{r}{nc}e^\frac{1}{n}\right)\right)}\right)r>0\\
 	\frac{1}{n^2W\left(\frac{r}{nc}e^\frac{1}{n}\right)}&\left(\frac{n^2+n+1-nW\left(\frac{r}{nc}e^\frac{1}{n}\right)}{n\left(1+W\left(\frac{r}{nc}e^\frac{1}{n}\right)\right)}\right)<\frac{c}{r}
 \end{align*}
 Since, $c/r>0$, we have
 $$	\frac{1}{n^2W\left(\frac{r}{nc}e^\frac{1}{n}\right)}\left(\frac{n^2+n+1-nW\left(\frac{r}{nc}e^\frac{1}{n}\right)}{n\left(1+W\left(\frac{r}{nc}e^\frac{1}{n}\right)\right)}\right)<0$$
 $$\iff n^2+n+1-nW\left(\frac{r}{nc}e^\frac{1}{n}\right)<0. $$

%
\subsection{Asymmetric case: Equilibrium Contracts for $n>2$}
\label{app:AsymmetricMultipleCPs}
\begin{thm} \label{cor:eql-NN}
	In the non-neutral regime equilibrium contract for $\CPi$ is given by 
	\begin{equation}
	\beta_i^{NN}=\begin{cases}
	0  &\mbox{  if }  \frac{r_i}{c} < 1, \\
	\frac{1}{W\left(\frac{r_i}{c}e\right)}  &\mbox{  if }  \frac{r_i}{c} \geq 1,
	\end{cases}
	\end{equation}
	Further, the effort levels of the ISP for are given by
	\begin{equation} \label{eqn:NN-a}
	a_i^{NN}=\max\left(\frac{\beta_i^{NN} r_i}{c}-1,0\right) \hspace{1.5mm} \forall i=1,2,\cdots n.
	\end{equation}
\end{thm}

\begin{thm} \label{thm:N-asy:gen}
	In the neutral regime, each $\CPi, i=1, 2,\cdots,n$ shares a positive fraction of the revenue at equilibrium with the ISP only if $r_i/c
	>1 \forall i=1,2,\cdots,n$ and $r_1,r_2,...,r_n$ are close enough to each other. Specifically, the equilibrium contract is as follows
	\begin{equation}
	\beta_i^N=\frac{\sum_{j=1}^{n}r_j}{n^2r_i
		W\left(\frac{\sum_{j=1}^{n}r_j}{n^2c}e^{\frac{1}{n}}\right)}-\frac{\sum_{j=1,j\ne i}^{n}r_j-(n-1)r_i}{nr_i} ;\forall i=1,2,...,n.
	\end{equation}
	and the equilibrium effort is $a^N=\left(\frac{\sum_{j=1}^{n}\beta_j r_j}{nc}-1\right)$.\\
	When $r_1>>r_2$, only CP$_1$ shares positive fraction at equilibrium, and the equilibrium contract is given as follows:
	\begin{equation}
	\beta_1^N=\frac{1}{W\left(\frac{r_1}{nc}e\right)}, \& \beta_i^*=0, \forall i=2,3,...,n
	\end{equation}
	and the equilibrium effort is $a^N=\left(\frac{\beta_1 r_1}{nc}-1\right) $
\end{thm}

{Proof:} Substituting the best action of ISP determined in CP$_i$'s optimization problem, we get:
$$\max_{\beta_i} \hspace{3mm} (1-\beta_i)r_i\log\left(\max\left(\frac{\sum_{j=1}^{n}\beta_jr_j}{nc},1\right)\right)$$
First order necessary condition for CP$_i$ gives
\begin{equation}
\frac{(1-\beta_i)r_i}{\sum_{j=1}^{n}\beta_jr_j}-\log \left(\frac{\sum_{j=1}^{n}\beta_jr_j}{nc}\right)=0 \hspace{1mm} \forall \hspace{1mm} i=1,2,...,n
\end{equation}
Comparing these set of eqns, we get,
$$(1-\beta_1)r_1=(1-\beta_2)r_2=...=(1-\beta_n)r_n$$
$$\implies \beta_ir_i=\beta_j r_j+ r_i-r_j; \hspace{1mm} \forall \hspace{1mm} i=1,2,...,n; i\ne j$$
$$\therefore\hspace{1mm} \sum_{j=1}^{n}\beta_jr_j= n\beta_ir_i+\sum_{j=1,j\ne i}^{n}r_j-(n-1)r_i \forall \hspace{1mm} i=1,2,...,n $$
Substituting, we get
\begin{align} \label{eqn:betan} \nonumber
&\frac{(1-\beta_i)r_i}{n\beta_ir_i+\sum_{j=1,j\ne i}^{n}r_j-(n-1)r_i}\\
&=\log \left(\frac{n\beta_ir_i+\sum_{j=1,j\ne i}^{n}r_j-(n-1)r_i}{nc}\right)
\end{align}
Adding $1/n$ to both the sides of eqn.(\ref{eqn:betan}), we get
\begin{align*}
&\frac{\sum_{j=1}^{n}r_j}{n\left(n\beta_ir_i+\sum_{j=1,j\ne i}^{n}r_j-(n-1)r_i\right)}\\
&=\log \left(\frac{n\beta_ir_i+\sum_{j=1,j\ne i}^{n}r_j-(n-1)r_i}{nc}\right)+\log e^{\frac{1}{n}}
\end{align*}
Rearranging and solving, we get
\begin{equation} \nonumber
{\beta}_i=\frac{\sum_{j=1}^{n}r_j}{n^2r_i
	W\left(\frac{\sum_{j=1}^{n}r_j}{n^2c}e^{\frac{1}{n}}\right)}-\frac{\sum_{j=1,j\ne i}^{n}r_j-(n-1)r_i}{nr_i} ; i=1,2,...,n
\end{equation}
Since, $r_1>r_2$, $\beta_1>0$, however, $\beta_2, \beta_3,...,\beta_n$ in above expression can tale negative value. therefore, the above solution holds only if $r_i$'s are sufficiently close s.t.  above solution is positive $\forall i=1,2,...,n$. Else, $\beta_2, \beta_3,...,\beta_n=0$, and $\beta_1$ is obtained from $\frac{(1-\beta_1)}{\beta_1}-\log \left(\frac{\beta_1r_1}{nc}\right)=0$. Solution of which is $\beta_1=\frac{1}{W\left(\frac{r_1}{nc}e\right)}$. \hfill\IEEEQED

\subsection{Proof of Proposition \ref{prop:contracts}}
\label{app:AsymmetricConstracts}
{ \it Part 1:	}We know that there exist some $r_1< r_1^*$, for which $\beta_2^{NN}$ is positive given by
\begin{equation} \nonumber
\beta_2^{N}=\frac{r_1+r_2}{4
r_2W\left(\frac{r_1+r_2}{4c}e^{0.5}\right)}-\frac{r_1-r_2}{2r_2}
\end{equation}
Now, differentiating $\beta_2^{N}$ w.r.t $r_2$, we get:
$$
\frac{\partial \beta_2^{N}}{\partial r_2 }=\frac{1}{4\left(1+W\left(r_2\frac{r_1+r_2}{4c}e^{0.5}\right)\right)}-\frac{1}{2} \leq 0 \hspace{0.75mm}\forall \hspace{0.5mm} r_1 \geq r_2
$$
which implies decreases$\beta_2^{N}$ with increase in $r_1$.\\
And for $r_1> r_1^*$, $\beta_2^{N}=0$. And $
\beta_2^{NN}=\frac{1}{W(\frac{r_2}{c}e^{0.5})}>0
$
which remain unchanged with increase in $r_1$. Also at $r_1 =r_2$ (symmetric case), $\beta_2^{NN}\geq \beta_2^{N}$. Thus, $\beta_2^{NN}\geq \beta_2^{N}\hspace{0.75mm}\forall \hspace{0.5mm} r_1 \geq r_2$.\\
{\it Part 2:} It is clear from the expression of $\beta_1^{NN}$ that it is decreasing in $r_1$
Now, there exist some $r_1> r_1^*$ for which $\beta_1^{N}$ is $\beta_1^{N}=\frac{1}{W(\frac{r_1}{2c}e)}$,
which also decreases with increase in $r_1$. Also, for $r_1 > r_1^*$, $\beta_1^{N}>\beta_1^{NN}$.
Now, consider the case when $r_1 \leq r_1^*$ where
\begin{equation} \nonumber
\beta_1^{N}=\frac{r_1+r_2}{4
	r_1 W\left(\frac{r_1+r_2}{4c}e^{0.5}\right)}-\frac{r_2-r_1}{2r_1 }
\end{equation}
Now, differentiating $\beta_1^{N}$ w.r.t $r_1$, we get:
$$
\frac{\partial \beta_1^{N}}{\partial r_1 }=\frac{\left(1+W\left(\frac{r_1+r_2}{4c}e^{0.5}\right)\right)\left(2W\left(\frac{r_1+r_2}{4c}e^{0.5}\right)-1\right)-r_1}{4r_1^2 W\left(1+W\left(\frac{r_1+r_2}{4c}e^{0.5}\right)\right)}$$
And it can take both negative and positive values depending upon value of $r_2$. Therefore, it if not apparent that whether $\beta_1^N$ increases or decreases when $r_1<r_1^*$. \hfill\IEEEQED 

\subsection{Proof of Proposition \ref{prop:CPUtility}} 
\label{app:AsymmetricCPUtility}
\noindent
{\it Part 1:}
Utility of $CP_1$ in non-neutral regime is given by $$U_{CP_1}^{NN}=(1-\beta_1^{NN})r_1\log\left(\frac{\beta_1^{NN}r_1}{c}\right)=\frac{(1-\beta_1^{NN})^2}{\beta_1^{NN}}r_1$$
(Using first order condition $\frac{(1-\beta_1^{NN})}{\beta_1^{NN}}=\log\left(\frac{\beta_1^{NN}r_1}{c}\right)$)\\
Utility of $CP_1$ in neutral regime is given by for all $r_1>r_1^*$ 
$$U_{CP_1}^{N}=(1-\beta_1^{N})r_1\log\left(\frac{\beta_1^{N}r_1}{2c}\right)=\frac{(1-\beta_1^{N})^2}{\beta_1^{N}}r_1$$
(Using first order condition $\frac{(1-\beta_1^{N})}{\beta_1^{N}}=\log\left(\frac{\beta_1^{N}r_1}{2c}\right)$)\\
We know for all $r_1>r_1^*$, $\beta_1^{N}>\beta_1^{NN}$.
\begin{align*}
\implies (1-\beta_1^{N})^2<(1-\beta_1^{NN})^2\text{and}&\hspace{1mm} \frac{1}{\beta_1^{N}}<\frac{1}{\beta_1^{NN}}\\
\implies  \frac{(1-\beta_1^{N})^2}{\beta_1^{N}}r_1<\frac{(1-\beta_1^{NN})^2}{\beta_1^{NN}}r_1,&\text{thus,} \hspace{1mm} U_{CP_1}^{N}<U_{CP_1}^{NN}
\end{align*}
{\it Part 2:}
Utility of $CP_2$ in non-neutral regime is given by $$U_{CP_2}^{NN}=(1-\beta_2^{NN})r_2\log\left(\frac{\beta_2^{NN}r_2}{c}\right)=\frac{(1-\beta_2^{NN})^2}{\beta_2^{NN}}r_2$$
(Using first order condition $\frac{(1-\beta_2^{N})}{\beta_2^{N}}=\log\left(\frac{\beta_2^{N}r_1}{c}\right)$)\\
Utility of $CP_2$ in non-neutral regime is given by for $r_1\geq r_1^*$ $$U_{CP_2}^{N}=(1-\beta_2^{N})r_2\log\left(\frac{\beta_1^{N}r_1+\beta_2^{N}r_2}{2c}\right)=\frac{(1-\beta_1^{N})}{\beta_1^{N}}r_2$$
(Using first order condition $\frac{(1-\beta_1^{N})}{\beta_1^{N}}=\log\left(\frac{\beta_1^{N}r_1}{2c}\right)$)\\
\begin{align*}
\text{Now,} \hspace{2mm}U_{CP_2}^{NN} \leq U_{CP_2}^{N} \hspace{2mm}&\text{iff}\hspace{2mm}\frac{(1-\beta_2^{NN})^2}{\beta_2^{NN}}\leq \frac{(1-\beta_1^{N})}{\beta_1^{N}}
\end{align*}
We know that when $r_1>r_1^*$, $\beta_1^N$ decreases with increase in $r_1$. Thus, RHS of above inequality is increasing in $r_1$. However, $\beta_2^{NN}$ remain unchanged with increase in $r_1$, implying that LHS of the above inequality is constant. Therefore, there exist some $r_1$, beyond which the above inequality holds.  
We know that $U_{CP_2}^{NN}$ remain constant with increase in $r_1$. \hfill\IEEEQED 


\subsection{Proof of lemma \ref{lma:ISPutilitycomp}}
\label{app:AssymtricISPUtility}
Utility of $ISP$ in non-neutral regime is given by
\begin{align*}
U_{ISP}^{NN}&=\beta_1^{NN}r_1\log\left(\frac{\beta_1^{NN}r_1}{c}\right)+\beta_2^{NN}r_2\log\left(\frac{\beta_2^{NN}r_2}{c}\right)\\ 
&\hspace{3mm}-c\left(\frac{\beta_1^{NN}r_1}{c}+\frac{\beta_2^{NN}r_2}{c}-2\right)\\
&=(1-2\beta_1^{NN})r_1+(1-2\beta_2^{NN})r_2+2c 
\end{align*}
(Using first order condition $\frac{(1-\beta_i^{N})}{\beta_i^{N}}=\log\left(\frac{\beta_i^{N}r_i}{c}\right); i=1,2$)\\
Utility of ISP in neutral regime is given by for $r_1\geq r_1^*$
\begin{align*}
U_{ISP}^{N}&=\beta_1^{N}r_1\log\left(\frac{\beta_1^{N}r_1}{2c}\right)-2c\left(\frac{\beta_1^{N}r_1}{2c}-1\right)\\
&=(1-2\beta_1^{N})r_1+2c
\end{align*}
(Using first order condition $\frac{(1-\beta_1^{N})}{\beta_1^{N}}=\log\left(\frac{\beta_1^{N}r_1}{2c}\right)$)\\ 
Now, $U_{ISP}^{NN}\geq U_{ISP}^{N}$ \textit{iff}
\begin{align*}
(1-2\beta_1^{NN})r_1+(1-2\beta_2^{NN})r_2&\geq (1-2\beta_1^{N})r_1\\
2(\beta_1^{N}-\beta_1^{NN})r_1&\geq -(1-2\beta_2^{NN})r_2\\
2 r_1 \left(\frac{1}{W(\frac{r_1}{2c}e)}-\frac{1}{W(\frac{r_1}{c}e)}\right)&\geq \left(2\frac{1}{W(\frac{r_2}{c}e)}-1\right)r_2
\end{align*}
LHS in increasing in $r_1$, however RHS remain unchanged. Therefore, there must exist some $r_1>r_1^*$ say $r_1^b$s.t. for all $r_1>r_1^b$ the above inequality holds. Also, plot shows that ISP is always better off in non-neutral regime. \hfill\IEEEQED 
%
%

%
\subsection{Proof of Lemma \ref{lma:totaleffortcomp}}
\label{app:AsymmtricTotalEffort}

\begin{align*}
	A^N\geq A^{NN} \iff \frac{\frac{r_1}{c}}{W(\frac{r_1}{2c}e)}-2 & \geq \frac{\frac{r_1}{c}}{W(\frac{r_1}{c}e)}+\frac{\frac{r_2}{c}}{W(\frac{r_2}{c}e)}-2\\
	\iff r_1 \left(\frac{1}{W(\frac{r_1}{2c}e)}-\frac{1}{W(\frac{r_1}{c}e)}\right)& \geq \frac{{r_2}}{W(\frac{r_2}{c}e)}
\end{align*}
LHS of above inequality is increasing in $r_1$ and RHS remains unchanged. With increase in $r_1$ for fixed $r_2$, the above inequality will start holding for some large value of $r_1$. Thus, the above inequality will hold for some large enough $r_1^a >>r_1^*$. \hfill\IEEEQED 